\documentclass[tradiabstract]{aa}
\usepackage{graphicx}
\usepackage{amsbsy}
\usepackage{natbib}
\usepackage{color}
\usepackage{txfonts}
\usepackage{url}
\newcommand{\vc}[1]{{\boldsymbol #1}}
\newcommand{\de}{\mathrm{d}}
\newcommand{\dpa}{\partial}

\newcommand{\nab}{\vc{\nabla}}

\DeclareMathSymbol{\varOmega}{\mathord}{letters}{"0A}
\DeclareMathSymbol{\varSigma}{\mathord}{letters}{"06}
\DeclareMathSymbol{\varPsi}{\mathord}{letters}{"09}
\newcommand{\Eq}[1]{Eq.~(\ref{#1})}

\newcommand{\Eqss}[2]{Eqs.~(\ref{#1})-(\ref{#2})}

\newcommand{\Fig}[1]{Fig.~\ref{#1}}
\newcommand{\Figs}[2]{Figs.~\ref{#1} and \ref{#2}}
\newcommand{\Tab}[1]{Table \ref{#1}}
\newcommand{\bl}{{\textcolor{white}{1}}}
\begin{document}

\title{High accretion rates in magnetised Keplerian discs\\mediated by a Parker
instability driven dynamo}
\titlerunning{High accretion rates in magnetised Keplerian discs}

\author{Anders Johansen\inst{1,2} \and Yuri Levin\inst{1}}
\authorrunning{Johansen and Levin} 

\offprints{\\A.\ Johansen (\email{ajohan@strw.leidenuniv.nl})}

\institute{Leiden Observatory, Leiden University, P.O.\ Box 9513, 2300 RA
Leiden, The Netherlands
    \and
    work partially done at Max-Planck-Institut f\"ur Astronomie, K\"onigstuhl
    17, 69117 Heidelberg, Germany}

\abstract{
Hydromagnetic stresses in accretion discs have been the subject of intense
theoretical research over the past one and a half decades. Most of the disc
simulations have assumed a small initial magnetic field and studied the
turbulence that arises from the magnetorotational instability. However,
gaseous discs in galactic nuclei and in some binary systems are likely to have
significant initial magnetisation. Motivated by this, we performed ideal
magnetohydrodynamic simulations of strongly magnetised, vertically stratified
discs in a Keplerian potential. Our initial equilibrium configuration, which
has an azimuthal magnetic field in equipartion with thermal pressure, is
unstable to the Parker instability. This leads to the expelling of magnetic
field arcs, anchored in the midplane of the disc, to around five scale heights
from the midplane. Transition to turbulence happens primarily through
magnetorotational instability in the resulting vertical fields, although
magnetorotational shear instability in the unperturbed azimuthal field plays a
significant role as well, especially in the midplane where buoyancy is weak.
High magnetic and hydrodynamical stresses arise, yielding an effective
$\alpha$-value of around 0.1 in our highest resolution run. Azimuthal magnetic
field expelled by magnetic buoyancy from the disc is continuously replenished
by the stretching of a radial field created as gas parcels slide in the linear
gravity field along inclined magnetic field lines. This dynamo process, where
the bending of field lines by the Parker instability leads to re-creation of
the azimuthal field, implies that highly magnetised discs are astrophysically
viable and that they have high accretion rates.
\keywords{Accretion, accretion disks -- Galaxy: center -- Instabilities --
MHD -- Turbulence}
}
\maketitle
\section{Introduction}

Since the seminal work of \cite{BalbusHawley1991} it has been widely recognised
that hydromagnetic stresses play a central role in the dynamics of Keplerian
gaseous discs. Much of the subsequent theoretical work has been devoted to the
study of the magnetic dynamo driven by a combination of magnetorotational
instability (MRI), rotation, and Keplerian shear \citep[][see
\citealp{BrandenburgSubramanian2005}, for an excellent review of astrophysical
dynamo theory]{Brandenburg+etal1995,Hawley+etal1996}. Typically, numerical
simulations of the MRI-driven dynamo begin with an initial zero net flux
magnetic field with an associated pressure which is a small fraction of the
thermal pressure. The idea is that the MRI-driven turbulence should increase
the characteristic coherence length of the magnetic field, and grow its mean
strength to a significant fraction of the equipartition value. The simulations,
however, have had at best a mixed success in explaining high and persistent
inflow rates observed in astrophysical accretion discs. Firstly, a number of
recent numerical experiments have shown that the effectiveness of the dynamo
depends in a critical way on the value magnetic Prandtl number, defined as the
ratio of the collisional kinematic viscosity to the magnetic diffusivity
\citep[][following suggestive non-disc simulations of
\citealp{Schekochihin+etal2005} and a careful study of previously published
MRI-turbulence results by
\citealp{Pessah+etal2007}]{LesurLongaretti2007,Fromang+etal2007}.
The upshot of this work is that when the Prandtl number is significantly less
than one, as is expected in most accretion discs \citep{BalbusHenri2008}, the
dynamo seems to fail\footnote{Also the magnitude of the turbulent magnetic
stresses decrease as the grid resolution increases (decreasing the numerical
diffusivity accordingly), with no convergence in sight
\citep{FromangPapaloizou2007}. In may well be,
however, that this problem is related to an insufficient size of the simulation
domain, and that bigger, stratified shearing boxes will show better convergence
\citep{RegevUmurhan2008}.}. Secondly, in the zero net flux simulations where
all parameters are chosen so that the dynamo works, the measured value of the
Shakura-Sunyaev parameter $\alpha$ is typically of order $10^{-3}$ and at most
$10^{-2}$
\citep[e.g.][]{Brandenburg+etal1995,Sano+etal2004,JohansenKlahr2005,FromangNelson2006}.
This is $1$--$2$ orders of magnitude smaller than what is required to explain
the high accretion rates in dwarf novae systems \citep{King+etal2007}.

The weak initial magnetic fields assumed in almost all disc-MRI simulations may
be unrealistically small for many astrophysical discs. We give here two
examples:\newline
\hskip .1in
1. Extended accretion discs around central supermassive black holes \citep[such
as the ones traced by megamasers in nearby galaxies; see
e.g.][]{Greenhill2007,Vlemmings+etal2007} are fed by molecular material in the
interstellar medium. Molecular clouds are known to have large scale
superthermal magnetic fields, and thus the initial magnetic fields in AGN discs
are likely to be comparable to or larger than the thermal equipartition values.
The $\sim$2 pc molecular circumnuclear disc in our galactic centre is permeated
by large scale equipartition magnetic fields
\citep{WardleKonigl1990,Hildebrand+etal1990} which are certain to play an
important dynamical role in its subsequent evolution.\newline \hskip .1in 2.
Accretion discs in close binary systems, which are fed by a Roche-lobe overflow
from a tidally distorted low mass star, may be initially magnetised. The
magnetisation $1/\beta$ of the gas as it departs from the donor star can be
estimated as
\begin{equation}\label{binary1}
  {1\over \beta}\sim {B_\star^2 \over 8\pi \rho c_s^2},
\end{equation}
where $B_\star$ (measured in Gauss) is the stellar magnetic field at the
surface, $c_s$ is the speed of sound at the stellar surface, and $\rho$ is the
density of gas as it becomes detached from the star. The density $\rho$ can be
estimated as
\begin{equation}\label{binary2}
  \rho\sim {\dot{M}\over 4\pi R_\star^2 c_s},
\end{equation}
where $\dot{M}$ is the average accretion rate of the companion, and $R_\star$
is the radius of the star. We get 
\begin{equation}
  {1\over \beta}\sim 0.3 \left({\dot{M}\over 10^{-9}M_{\odot}/\hbox{yr}}\right)^{-1}
               \left({c_s\over 3\hbox{km/s}}\right)^{-1}\left({B_\star \over 1\hbox{G}}\right)^2
               \left({R_\star\over 10^{11}\hbox{cm}}\right)^2.
\label{binary3}
\end{equation}
The streaming and shearing motion after the gas detaches from the star can
further amplify the magnetic field. Clearly, there is a realistic parameter
range where the initial magnetisation is high. 

Motivated by these astrophysical considerations, in this paper we perform
numerical experiments on gas discs which contain initially strong magnetic
fields, with magnetic pressure comparable to that of the  gas. Specifically, we
initialise the azimuthal field so that it is subject to the Parker instability
\citep[PI,][]{Parker1966} in the vertical stratification.

In terms of initial conditions our physical set up is close to
\cite*{Machida+etal2000}. However, \cite{Machida+etal2000} focused on the
formation and evolution of a disc corona, and the spirit or their numerical
experiment was different from ours. They have simulated the whole circular
disc, and have introduced reflecting boundary conditions at the midplane of
the disc, which suppresses field anchoring in the midplane. By contrast, our
purpose is to understand the long-term dynamics of the disc, possible field
confinement, and the dynamo processes related to strong magnetic fields.
Therefore, our shearing-box simulations focus on a small part of the disc and
study it with high numerical resolution. We simulate the fluid both below and
above the midplane, and thus have no artificial boundary conditions at the
midplane of the disc.

We find that both the short term and, more importantly, the long term behaviour
of initially strongly magnetised discs is radically different from that of
their weakly magnetised counterparts. We observe the following three-step
dynamics: (a) the Parker instability expels azimuthal field in huge arcs,
creating vertical field which becomes the seed for a strong magnetorotational
instability, (b) matter sliding down inclined field lines stretches the
azimuthal
magnetic field and creates a vertically dependent large scale mean radial
field, and (c) the Keplerian shear recreates azimuthal field from the
stretching of the radial field. The latter step closes the dynamo cycle, in
much the same way as was sketched by \cite{ToutPringle1992} a decade and a half
ago, although non-axisymmetric magnetorotational instability in the
azimuthal field also plays an important role in creating accretion stresses in
our simulations
\citep{BalbusHawley1992,FoglizzoTagger1995,TerquemPapaloizou1996}. The
azimuthal field remains strongly concentrated towards the disc midplane; this
is in contrast with the simulations of Parker instability which do not include
strong Keplerian shear in the radial direction \citep{Kim+etal1998}. We show
that the dynamo is robust and stable over at least tens of orbital periods, and
that the accretion torque {\it increases} if we use a finer grid. We observe
$\alpha$-values as high as $0.1$ in our highest resolution run.


The plan of the paper is as follows. In \S\ref{s:numerics} we detail the
mechanics of our numerical experiments, and in the following section
(\S\ref{s:noshear}) we test our computer code by comparing the results of
Parker instability simulations without Keplerian shear to the extensive
literature that exists on the PI under rigid rotation. In \S\ref{s:shear} we
perform simulations with the Keplerian shear included, and present our main
results. We devote the next section (\S\ref{s:confinement}) to analysing the
confinement of azimuthal flux to the disc by a dynamo process. In
\S\ref{s:conclusions} we conclude with the discussion of the astrophysical
implications and possible future improvements of our work. 

\section{Description of the numerical experiment}
\label{s:numerics}

In this section we describe the numerical method we use to solve the equations
of ideal magnetohydrodynamics, the set up of an initial condition that is
unstable to the Parker instability, and technical issues such as dissipation
terms and boundary conditions. We use the Pencil
Code\footnote{See\\\url{http://www.nordita.dk/data/brandenb/pencil-code/}.} to
solve the relevant partial differential equations.

\subsection{Dynamical equations}

We consider a local corotating patch of a Keplerian accretion disc. The
coordinate axes are oriented such that $x$ points radially away from the
central gravity
source, $y$ points along the main orbital flow, while $z$ points vertically out
of the disc parallel to the Keplerian rotation vector $\vc{\varOmega}$. In the
shearing sheet approximation the equation of motion, for the velocity field
$\vc{u}$ relative to the Keplerian flow, is
\begin{eqnarray}\label{eq:eqmot}
  \frac{\dpa \vc{u}}{\dpa t} +
    (\vc{u} \cdot \nab) \vc{u} + u_y^{(0)} \frac{\dpa \vc{u}}{\dpa y} &=& 
    2 \varOmega u_y \vc{e}_x - (2-q) \varOmega u_x
    \vc{e}_y - \varOmega^2 z \vc{e}_z \nonumber \\
    & & \hspace{0.5cm} + \frac{1}{\rho^{*}} \vc{J} \times \vc{B} -
    \frac{1}{\rho} \nab P + \vc{f}_\nu(\vc{u}) \, .
\end{eqnarray}
Here $q=\dpa\ln\varOmega/\dpa\ln r$ is the shear parameter, with $q=0$ for
rigid rotation and $q=3/2$ for Keplerian rotation. The left hand side includes
advection both by the velocity field $\vc{u}$ itself and by the linearised
Keplerian flow $u_y^{(0)}=-q \varOmega x$. The first two terms on the
right hand side represent the Coriolis force in the $x$- and $y$-directions,
modified in the $y$-component by the radial advection of the Keplerian flow,
$\dot{u}_y=-u_x \dpa u_y^{(0)}/\dpa x$. The third term is the linearised
vertical component of the gravity of the central object, while the Lorentz
and pressure gradient forces appear in the usual way. The high order
numerical scheme of the Pencil Code has very little numerical dissipation from
time-stepping the advection term \citep{Brandenburg2003}, so we add explicit
viscosity through the term $\vc{f}_\nu(\vc{u})$, described in detail in
\S\ref{s:dissipation}

\subsubsection{Alfv\'en limiter}

In this paper we consider stratified models with 4--16 orders of magnitude of
range in densities. In low-density regions the Alfv\'en speed
\begin{equation}\label{eq:va}
  v_{\rm A} = \frac{B}{\sqrt{\mu_0 \rho}}
\end{equation}
may become very high when strong magnetic fields are transported away from
regions near the midplane to form a low density corona, and we must reduce
$v_{\rm A}$ artificially in order to avoid very low time-steps in the explicit
numerical scheme of the Pencil Code\footnote{In these magnetised, low-density
regions the field is almost force-free, and no dynamical Alfv\'en waves are
excited.}. We have done so by replacing the density $\rho$ in the Lorentz force
term of \Eq{eq:eqmot} by a modified density $\rho^{*}$ defined as
\begin{equation}\label{eq:rhostar}
  \frac{1}{\rho^*} = \frac{1}{\rho}
  \left[1+\left(\frac{v_{\rm A}^2}{M^2}\right)^n \right] ^{-1/n} \, .
\end{equation}
Here $M$ is a limiting value for the Alfv\'en speed and $n$ is an index that
smooths the transition from the regime where the Alfv\'en speed is not affected
($v_{\rm A}<M$) to the region where the limiter applies ($v_{\rm A}>M$). This
yields an effective Alfv\'en speed $v_{\rm A}^* \approx v_{\rm A}$ in the limit
$v_{\rm A}\ll M$, while in the limit of high Alfv\'en speeds the effective
speed is
\begin{equation}\label{eq:vastar}
  v_{\rm A}^* = \frac{B}{\sqrt{\mu_0 \rho^*}} \rightarrow \frac{B}{\sqrt{\mu_0
  \rho}} \frac{M}{v_{\rm A}} = M \quad \textrm{for} \quad v_{\rm A} \gg M \, .
\end{equation}
We used a smoothing index of $n=5$. We are grateful to T.\ Heinemann for
implementing this form of the Alfv\'en limiter in the Pencil Code
\citep{Heinemann+etal2007}. A similar
Alfv\'en speed limiter was used by \cite{MillerStone2000}. The Alfv\'en limiter
in the form presented here does not conserve momentum, but we have experimented
with $M^2=10$ and $M^2=100$ and found no qualitative differences in our
results. Thus we apply $M^2=100$ to all 2-D simulations, and $M^2=10$ to the
3-D simulations to get a longer time-step in those computationally expensive
runs.

\subsubsection{Induction equation}

The magnetic vector potential $\vc{A}$ is evolved through the induction equation
\begin{equation}\label{eq:ind}
  \frac{\dpa \vc{A}}{\dpa t} + u_y^{(0)} \frac{\dpa \vc{A}}{\dpa y} = \vc{u}
  \times \vc{B} + q \varOmega A_y \vc{e}_x + \vc{f}_\eta(\vc{A}) 
  \, .
\end{equation}
The explicit resistivity term $\vc{f}_\eta(\vc{A})$ is discussed in
\S\ref{s:dissipation}. Working with the vector potential $\vc{A}$ rather than
the magnetic field $\vc{B}$ has the advantage that $\vc{B}=\nab\times\vc{A}$ is
kept solenoidal ($\nab\cdot\vc{B}=0$) by mathematical construction. The
stretching of the radial component of the magnetic field by the Keplerian shear
is represented by the second term on the right hand side of \Eq{eq:ind}
[Brandenburg et al., 1995]. The current density $\vc{J}$, needed for
the Lorentz force in \Eq{eq:eqmot}, is calculated from Amp\`ere's law
$\vc{J}=\mu_0^{-1}\nab\times\vc{B}$. We set the vacuum permeability $\mu_0=1$.

\subsubsection{Continuity equation and equation of state}

The mass density is evolved in its logarithmic form
\begin{equation}\label{eq:cont}
  \frac{\dpa \ln \rho}{\dpa t} + \vc{u}\cdot\nab\ln\rho + u_y^{(0)}\frac{\dpa
  \ln \rho}{\dpa y} = - \nab \cdot \vc{u} + f_{\rm D}(\ln \rho) \, .
\end{equation}
The last term on the right hand side is an explicit diffusion term (see
\S\ref{s:dissipation}). It is advantageous to evolve the logarithmic mass
density when a large dynamical range in density is considered. Also, in the
initial magnetohydrostatic equilibrium, $\ln \rho$ varies parabolically with
$z$, and the Pencil Code's finite difference approach to spatial derivatives
yields a perfect derivative of parabolic curves \citep{Brandenburg2003}.
\begin{table*}[!t]
  \caption{Simulation parameters.}
  \begin{center}
    \begin{tabular}{lccccccc}
      \hline
      \hline
      Run & $L_x \times L_y \times L_z$
          & $N_x \times N_y \times N_z$
          & $\beta$ & $\langle B_z \rangle$ &
          $\varOmega$ & $q$ & $\Delta t$ \\
      \hline
      R2D\_256       & $\bl0.0\times24.0\times12.0$ &
          $\bl\bl1\times128\times256$ &  $1.0$ & $0.0$ &
          1.0 & 0  & 100\\
      R2D\_512       & $\bl0.0\times24.0\times12.0$ &
          $\bl\bl1\times256\times512$ &  $1.0$ & $0.0$ &
          1.0 & 0  & 100\\
      R2D\_256\_Lz24 & $\bl0.0\times24.0\times24.0$ &
          $\bl\bl1\times256\times512$ &  $1.0$ & $0.0$ &
          1.0 & 0  & 100\\
      R3D\_256       & $12.0\times24.0\times12.0$ &
          $128\times128\times256$ &  $1.0$ & $0.0$ &
          1.0 & 0  & \bl30\\
      S3D\_256       & $12.0\times24.0\times12.0$ &
          $128\times256\times128$ &  $1.0$ & $0.0$ &
          1.0 & 3/2 & \bl30\\
      S3D\_512       & $12.0\times24.0\times12.0$ &
          $256\times512\times256$ &  $1.0$ & $0.0$ &
          1.0 & 3/2 & \bl20\\
      S3D\_256\_b3   & $12.0\times24.0\times12.0$ &
          $128\times256\times128$ &  $3.0$ & $0.0$ &
          1.0 & 3/2 & \bl30\\
      S3D\_256\_Lz18 & $12.0\times24.0\times18.0$ &
          $128\times256\times192$ &  $1.0$ & $0.0$ &
          1.0 & 3/2 & \bl20\\
      S3D\_256\_Bz0.03 & $12.0\times24.0\times18.0$ &
          $128\times256\times192$ &  $1.0$ & $0.03$ &
          1.0 & 3/2 & \bl20\\
      \hline
    \end{tabular}
  \end{center}
  The first column gives the name of the simulation, while the box size, in
  units of the thermal scale height $H=c_{\rm s} \varOmega^{-1}$, and the grid
  resolution are given in the following two columns. The parameters given in
  the last five columns are: the ratio of thermal to magnetic pressure $\beta$
  in the initial azimuthal field, the mean imposed vertical field $\langle
  B_z\rangle$ in units of the midplane azimuthal field, the angular frequency
  $\varOmega$ of the box, the shear parameter $q$, and finally the simulation
  time in orbits.
  \label{t:parameters}
\end{table*}

We close the dynamical equation system by applying an isothermal equation of
state with $P=c_{\rm s}^2 \rho$. The sound speed $c_{\rm s}$ is assumed
constant.

\subsection{Dissipation terms}\label{s:dissipation}

The viscosity term $\vc{f}_\nu$ of \Eq{eq:eqmot} appears in full generality as
\begin{eqnarray}\label{eq:viscosity}
  \vc{f}_\nu &=& \nu_3
  \left[\nabla^6\vc{u}+(\vc{S}^{(3)}\cdot\nab\ln\rho)\right] \nonumber \\
  & & + \nu_{\rm shock} \left[
  \nab\nab\cdot\vc{u}+(\nab\cdot\vc{u})(\nab\ln\rho)\right]
  + \nab\nu_{\rm shock}(\nab\cdot\vc{u}) \, .
\end{eqnarray}
The viscosity includes both hyperviscosity with coefficient $\nu_3$ and shock
viscosity with coefficient $\nu_{\rm shock}$. A simplified third order
rate-of-strain tensor $\vc{S}^{(3)}$ is here defined as
\begin{equation}
  S_{ij}^{(3)} = \frac{\dpa^5 u_i}{\dpa x_j^5} \, .
\end{equation}
The shock viscosity coefficient is obtained by taking the negative part of the
divergence of $\vc{u}$, then taking the maximum over three grid cells in each
direction, and finally smoothing over three grid cells, to obtain
\begin{equation}\label{eq:nu_shock}
  \nu_{\rm shock}=c_{\rm shock} \langle {\rm max}[-\nab\cdot\vc{u}]_+ \rangle
    (\delta x)^2 \, .
\end{equation}
Here $c_{\rm shock}$ is a parameter of order unity \citep{Haugen+etal2004}.

For the resistivity term $\vc{f}_\eta$ in \Eq{eq:ind} we include both
hyperresistivity and shock resistivity,
\begin{eqnarray}
  \vc{f}_\eta(\vc{A}) &=& \eta_3 \nabla^6 \vc{A} + \eta_{\rm shock}
  \nabla^2\vc{A}+\nab \eta_{\rm shock}\nab\cdot\vc{A} \, .
\end{eqnarray}
This form is the resistivity conserves all components of the magnetic field
$\vc{B}$. We set $\eta_{\rm shock}=\nu_{\rm shock}$.

Mass diffusion consists of shock diffusion,
\begin{eqnarray}
  \vc{f}_D(\ln\rho) &=& D_{\rm shock} [\nabla^2\ln\rho+(\nab\ln\rho)^2] + \nab
  D_{\rm shock}\cdot\nab\ln\rho \, ,
\end{eqnarray}
where $D_{\rm shock}=\nu_{\rm shock}$ is defined in \Eq{eq:nu_shock}. We also
upwind the finite differencing of the advection term in the continuity equation
\citep{Dobler+etal2006}. Mass diffusion and upwinding are necessary to damp out
spurious small scale modes that are left behind due to the dispersion error in
the finite differencing of the advection term in the continuity equation.

We set shock diffusivity coefficients to $c_{\rm shock}=\eta_{\rm shock}=D_{\rm
shock}=6.0$ in all simulations to dissipate energy in shocks forming far away
from the midplane of the disc. In run S3D\_256\_Lz18, which is extended
vertically compared to the other 3-D simulations, we had to double $c_{\rm
shock}$ to dissipate enough energy in the regions above six scale heights from
the midplane.

We test the dependence of our results on dissipation parameters by keeping the
mesh Reynolds number approximately constant with increasing resolution, i.e.\
scaling hyperdissipation parameters by $(\delta x)^5$ and shock dissipation
parameters by $(\delta x)^2$. Higher resolution simulations thus probe the
effect of decreasing the dissipation coefficients.

\subsection{Initial conditions}
\label{s:ic}

We assume that the initial magnetic field is toroidal, a reasonable assumption
since Keplerian shear efficiently generates coherent toroidal field from a
small radial component. Furthermore, we tune the initial field so that (a)
magnetic pressure is the constant fraction $\beta^{-1}$ of the gas pressure,
and (b) vertical hydrostatic equilibrium is enforced, i.e.
\begin{equation}\label{eq:maghydeq}
  0 = - \rho(z) \varOmega^2 z - {\partial\over \partial z}\left(P+\frac{B_y^2}{2\mu_0}\right)
    = - \rho(z) \varOmega^2 z - (1+\beta^{-1}) \frac{\partial P}{\partial z}\, .
\end{equation}
Mathematically, the initial density, magnetic field, and vector potential are
given by
\begin{eqnarray}
  \rho(z) &=& \rho_0 \exp[-z^2/(2 H_\beta^2)] \, , \\
  \frac{B_y^2}{2\mu_0} &=& \beta^{-1} c_{\rm s}^2 \rho_0 \exp[-z^2/(2
  H_\beta^2)] \, , \label{eq:bymhse} \\
  A_x(z) &=& \sqrt{2 \pi \mu_0 \beta^{-1} c_{\rm s}^2 \rho_0}
           H_\beta {\rm erf}[z/(2 H_\beta)] \, .
\end{eqnarray}
Here $H_\beta=\sqrt{1+\beta^{-1}} c_{\rm s}/\varOmega$ is the gas scale height,
and $\rho_0$ is the initial density in the midplane. The thermal pressure
alone would give rise to a scale height of $H=H_\infty=c_{\rm s}/\varOmega$. We
will use this more familiar scale height as our unit of length, setting $H=1$
throughout the paper.

Since we shall consider many (six to twelve) scale heights above and below the
midplane, we run into the problem that the finite differencing of the magnetic
pressure gradient underestimates its actual value and leads to spurious
accelerations far away from the midplane. This is not a problem for the
equilibrium density stratification, since the logarithmic density varies as a
parabola with height over the midplane, and any parabolic shape has a perfect
numerical derivative in the symmetric finite difference scheme of the Pencil
Code. In order to give magnetic stratification equal opportunity as the
density, and to obtain a perfect initial magnetohydrostatic equilibrium, we
measure magnetic field and current density relative to their initial values, as
explained below.

We split the field and the current into two components, as follows. For the
Lorentz force in \Eq{eq:eqmot} and the induction term in \Eq{eq:ind} we replace
\begin{eqnarray}
  \vc{B} &\rightarrow& \vc{B}^{(0)} + B_y^{(1)}(z) \vc{e}_y \, , \\
  \vc{J} &\rightarrow& \vc{J}^{(0)} + J_x^{(1)}(z) \vc{e}_x \, .
\end{eqnarray}
The initial magnetic field $B_y^{\rm (0)}(z)$ is given by \Eq{eq:bymhse},
while the initial current density $J_x^{\rm (0)}(z)$ is calculated as
\begin{eqnarray}
  J_x^{(0)}(z) &=& -\frac{1}{\mu_0}\frac{\dpa B_y}{\dpa z} \nonumber \\
         &=& \sqrt{2\mu_0^{-1}\beta^{-1} c_{\rm s}^2 \rho_0} [z/(2 H_\beta^2)]
         \exp[-z^2/(4 H_\beta^2)] \, .
\end{eqnarray}
The effect of this splitting is that the initial, mean field is not exposed to
any (numerical or explicit) resistivity. Since we wish for resistive effects to
go to zero with increasing resolution\footnote{This is because our numerical
resistivity is always orders of magnitude greater than in realistic accretion
discs.}, the splitting does not introduce any extra spurious effects. We have
checked that the Parker instability develops similarly in models with and
without this field splitting, and have found that the models without splitting
indeed converge towards the models with splitting as the resolution increases.
\begin{figure*}
  \includegraphics[width=0.5\linewidth]{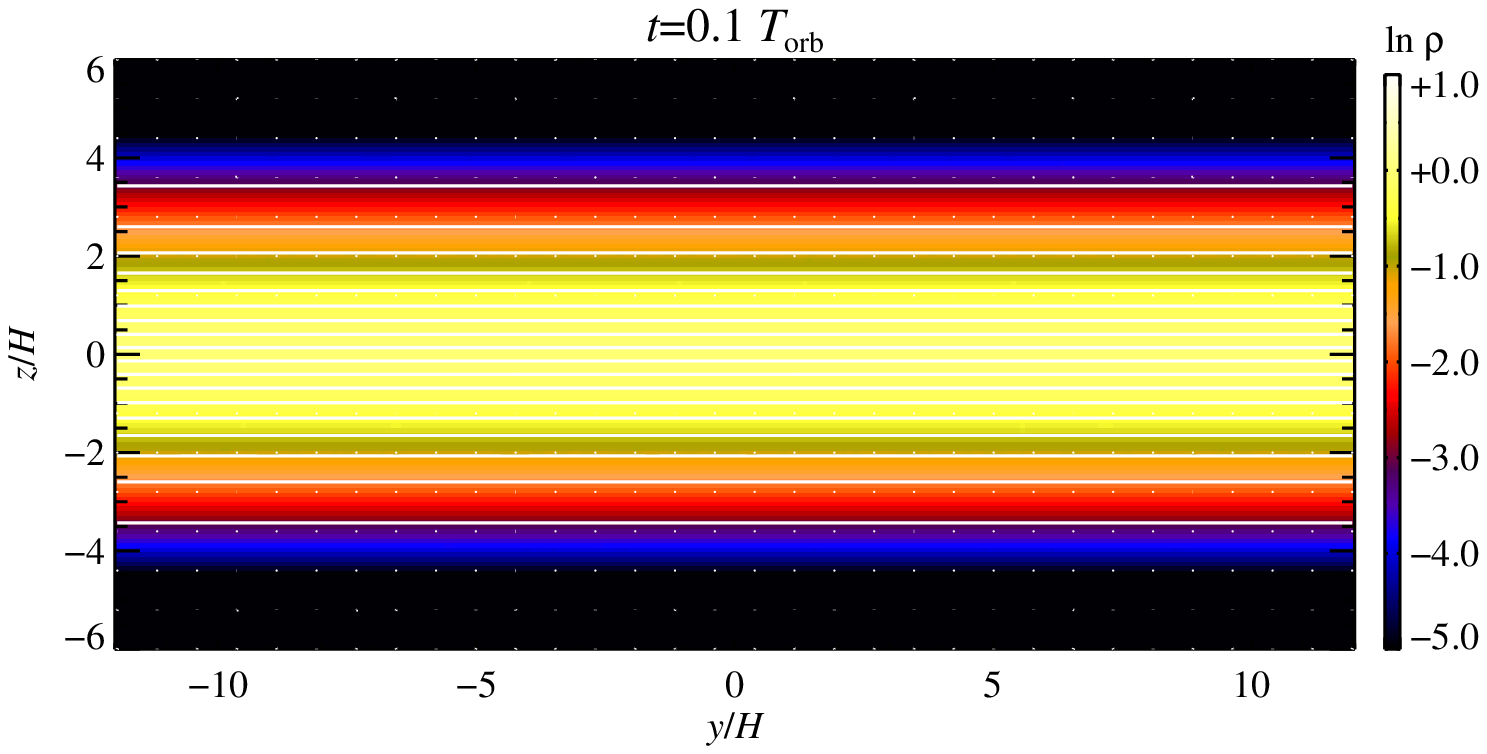}
  \includegraphics[width=0.5\linewidth]{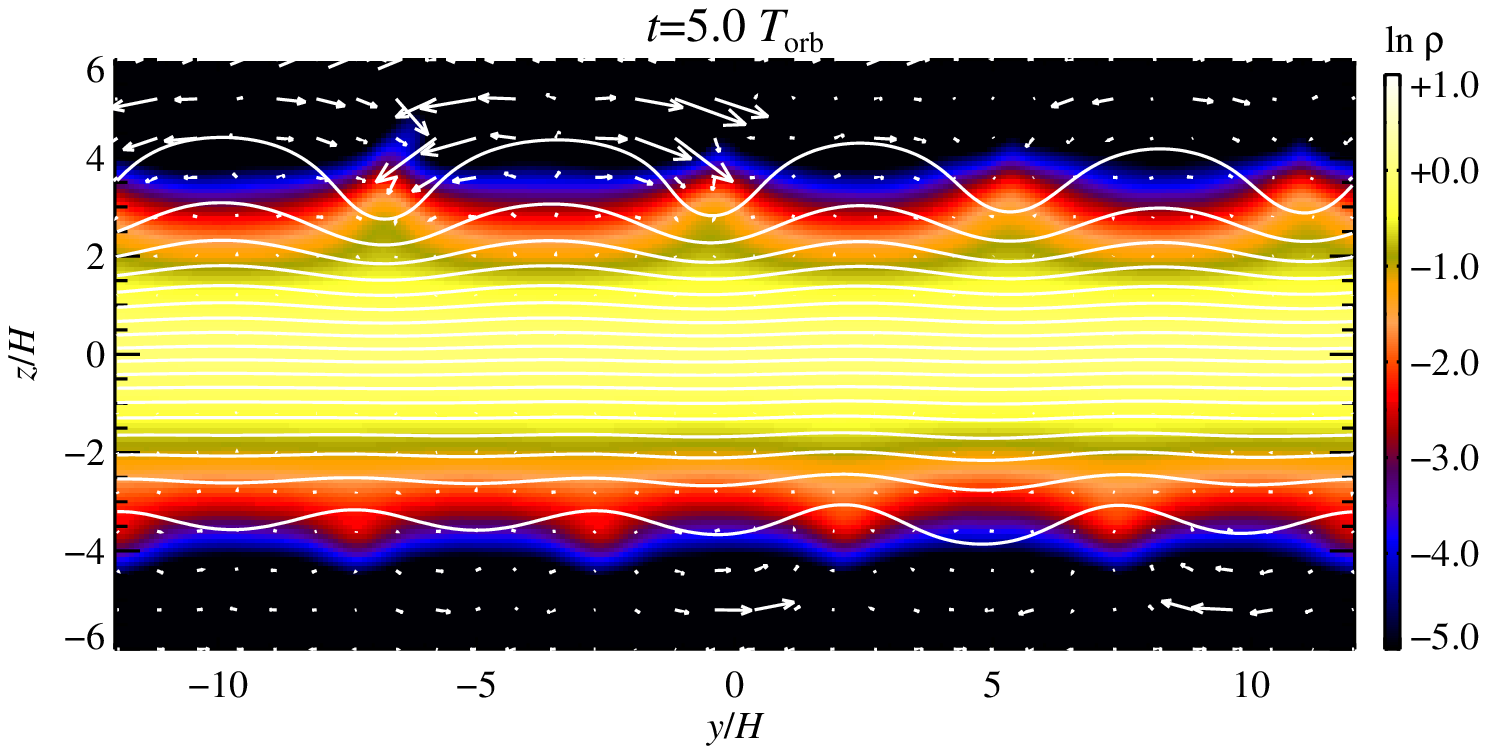}
  \includegraphics[width=0.5\linewidth]{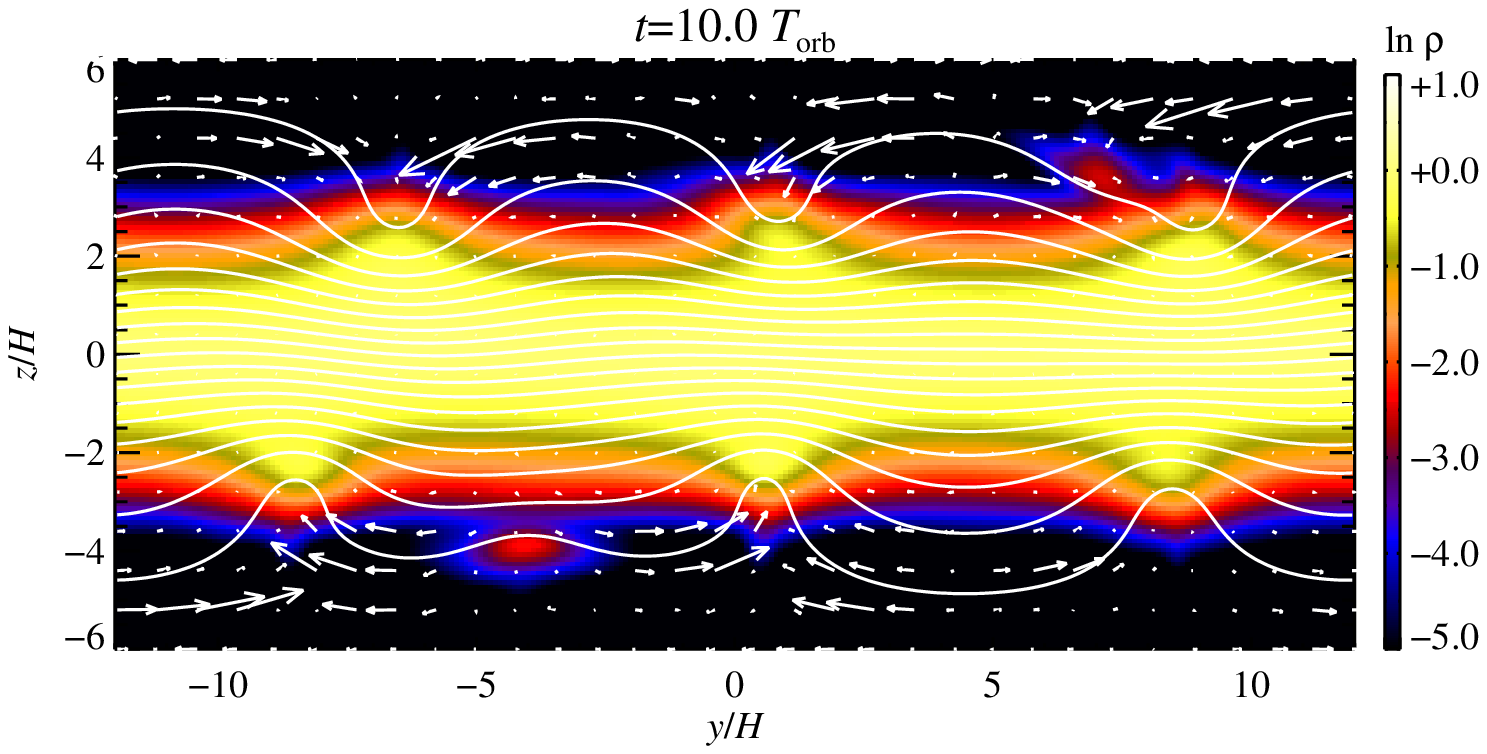}
  \includegraphics[width=0.5\linewidth]{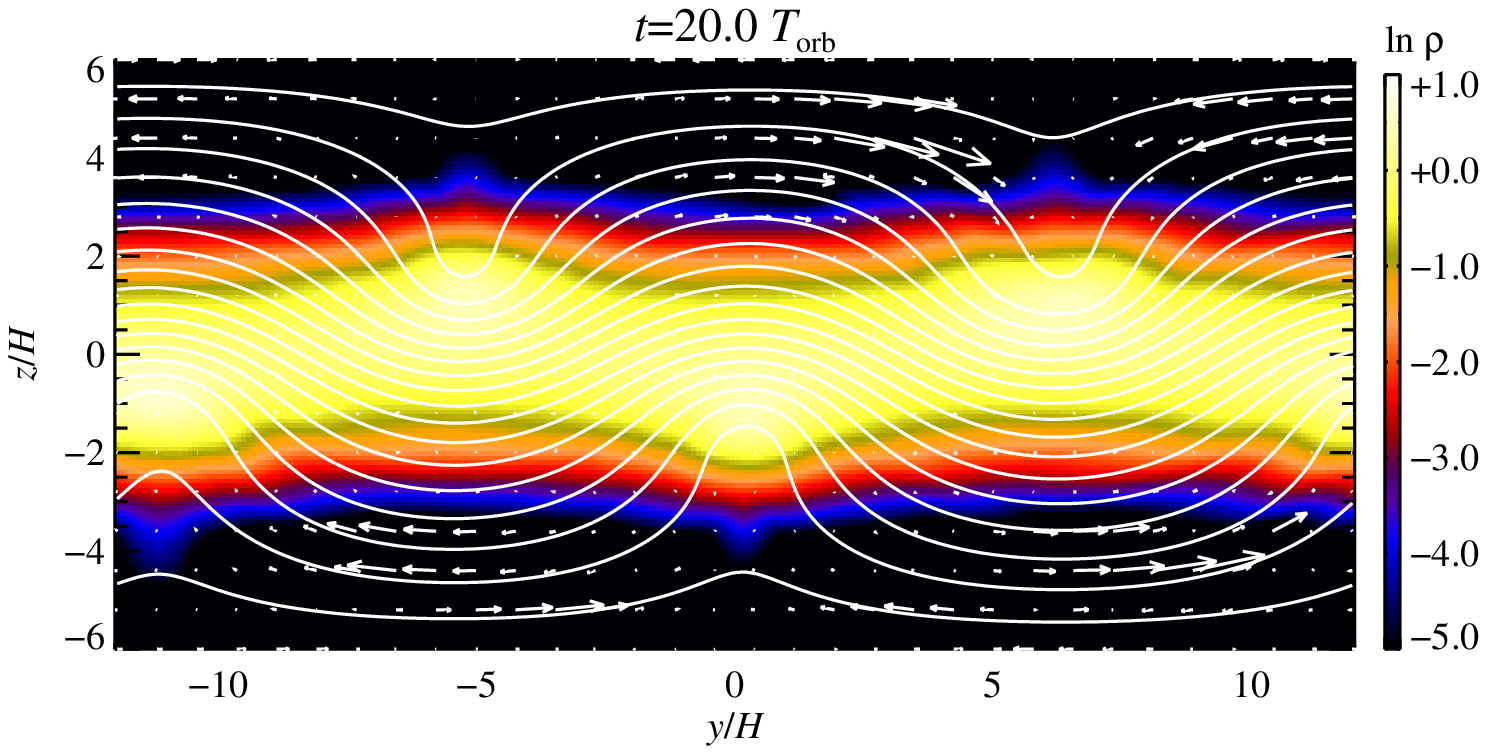}
  \caption{Evolution of the Parker instability in 2-D rigid rotation (run
    R2D\_256). Overlaid on the density are magnetic field streamlines (white
    lines) and velocity field vectors (white arrows, averaged over 8 grid
    points, the longest arrows represent approximately four times the sound
    speed). The initial stratification is unstable to magnetic buoyancy, and
    magnetic field arcs begin to rise from the midplane. The arcs merge to
    form longer arcs, and eventually the system settles down into a new
    equilibrium state with two superarcs and four dense pockets of matter in
    the midplane.}
  \label{f:lnrho_ax}
\end{figure*}

We perturb the initial Keplerian velocity field by Gaussian noise of
amplitude $\delta u \sim 10^{-3}$. This perturbation seeds the linear Parker
instability, and the noise also contains leading waves that can be amplified by
non-axisymmetric magnetorotational instability in the azimuthal magnetic field
in simulations with Keplerian shear.

\subsection{Boundary conditions}

We set the usual shearing sheet boundary conditions in the $x$ and $y$
directions (shear-periodic in $x$ and periodic in $y$). At the upper and lower
boundaries we impose free-slip conditions, $\dpa u_x/\dpa z=\dpa u_y/\dpa
z=u_z=0$, for the velocity field $\vc{u}$ and perfect conductor, $A_x=A_y=\dpa
A_z/\dpa z=0$, for the magnetic vector potential $\vc{A}$. This effectively
allows the radial and azimuthal components of the magnetic field to evolve
freely at the upper and lower boundary planes, while the vertical component of
the magnetic field is forced to be zero. Globally the average value of all
components of the magnetic field are then conserved.
Perfect conductor boundary
conditions do not allow azimuthal flux to flow out of the box, and therefore
there is a concern that the presence of the boundaries will artificially
enhance the field in the disc. To address this concern, we have made sure that
our results have converged with increasing vertical extent of the box; see
\S\ref{s:confinement} and \Figs{f:bymean_bzrms_z}{f:Bx_z_t}.

For the mass density $\ln \rho$ we set a symmetric boundary condition with
$\dpa\ln\rho/\dpa z=0$ at the upper and lower boundaries. Although this
precludes the gas pressure gradient from participating in magnetohydrostatic
equilibrium there, the free-slip boundary condition for the velocity field,
with zero vertical velocity component, means that this does not cause any
unwanted accelerations.

\subsection{Simulation parameters}

We perform a series of simulations of the evolution of Parker and
magnetorotational instabilities, usually starting from an equipartition
($\beta=1$) accretion disc in magnetohydrostatic equilibrium (as described in
\S\ref{s:ic}). Simulation parameters are written in \Tab{t:parameters}. The
main runs are the shearing sheet simulations S3D\_*, representing varying
resolution (S3D\_256, S3D\_512), higher initial ratio of gas to magnetic
pressure (S3D\_256\_b3), increased vertical extent of the box (S3D\_256\_Lz18) and
an exploration of the effect of a moderate amount of net vertical field
(S3D\_256\_Bz0.03). For testing purposes we also run a series of 2-D and 3-D
rigid rotation simulations (R2D\_256, R2D\_512, R2D\_256\_Lz24, R3D\_256),
which we present in the next section.

\section{Testing the code: evolution of the Parker instability under rigid
rotation}
\label{s:noshear}

In this section we study the evolution of the ``pure'' Parker instability in
2-D and 3-D simulations without shear \citep{Parker1966,Kim+etal1997}. This
allows us to test our code against the previously published results, and to
review previous work done on the evolution of the Parker instability in systems
with no systematic shear.

\cite{Basu+etal1997} performed 2-D inertial frame simulations of the non-linear
evolution of the Parker instability. They found that primarily antisymmetric
midplane crossing modes are excited by the random noise initial condition. The
instability develops quickest in the tenuous gas high over the midplane. As
the magnetic field lines rise in big arcs, matter streams unobstructed down the
lines that are no longer horizontal. Dense pockets of gas (approximately a
factor two times more dense than the average state) form in the midplane,
anchoring the rising magnetic field lines there. The final state of the Parker
instability, in which the tension in the bent magnetic field lines balances the
buoyancy, has a lower energy than the initial equilibrium state
\citep{Mouschovias1974} and is therefore preferred.
\begin{figure}
  \includegraphics{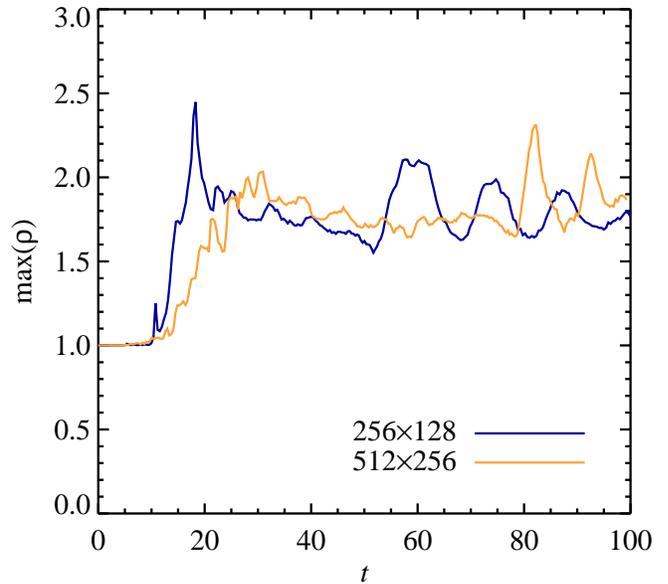}
  \caption{The maximum gas density as a function of time for 2-D simulations of
    the Parker instability (under rigid rotation). Increasing resolution leads
    to no significant change of the maximum density, showing that the Parker
    instability operates in a robust way already  at $256\times128$ grid
    points.}
  \label{f:rhomax_t_2D}
\end{figure}
\begin{figure}
  \includegraphics{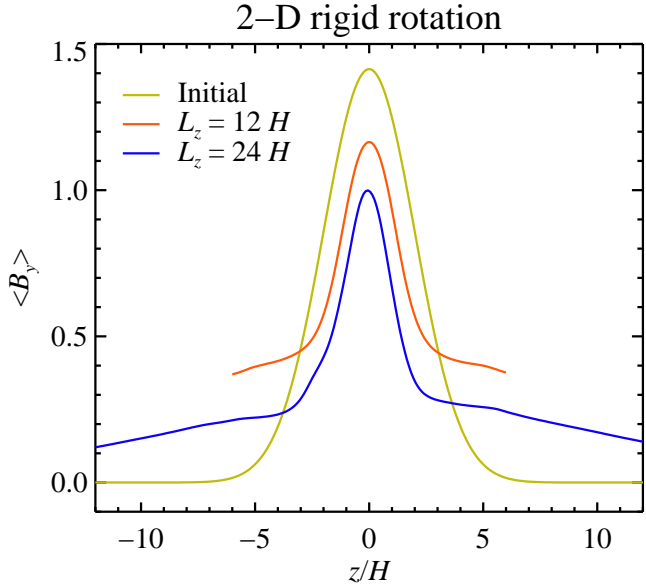}
  \caption{Mean azimuthal magnetic field, averaged over the $y$-direction, as a
    function of height $z$ over the midplane in 2-D simulations with rigid
    rotation. The initial Gaussian magnetic pressure
    profile is lifted by the Parker instability, reducing the azimuthal flux by
    a factor of approximately 1.5 in the midplane. The flux is transported to
    above four scale heights from the midplane, and an approximately force
    free corona develops at these high altitudes. Increasing the vertical
    extent of the computational domain from 12 to 24 scale heights causes the
    flux to be transported slightly further away from the midplane, but the
    overall shape of the magnetic field distribution stays approximately the
    same.}
  \label{f:Bymxy_z}
\end{figure}

In \Fig{f:lnrho_ax} we show the evolution of the Parker instability obtained
with the Pencil Code under rigid rotation in the 2-D $(y,z)$ plane (run
R2D\_256). Magnetic field lines initially rise in arcs of wavelength
$\lambda_{\rm PI}\approx 7 H$. As matter streams sideways along the rising
field lines (see arrows in \Fig{f:lnrho_ax}) dense clumps of gas appear in the
midplane after 10 orbits. Eventually the magnetic arcs merge and form two
superarcs with four dense gas clumps in the midplane.

In \Fig{f:rhomax_t_2D} we show the maximum gas density in the box as a function
of time. The clumps in the midplane are overdense by around factor $1.6-2.2$
compared to the initial state, and this contrast is relatively well converged
already at $256\times128$ grid points (as is the non-linear evolution of the
Parker instability in general). Our simulations are in good qualitative
agreement with the \cite{Basu+etal1997} simulations.

\subsection{Evolution of the azimuthal field}

In \Fig{f:Bymxy_z} we show the mean azimuthal magnetic field as a function of
height over the midplane in the saturated state of the 2-D Parker instability
under rigid rotation. The initial azimuthal field is redistributed by the
Parker instability so that some of the midplane flux ends above four scale
heights from the midplane. This scale is similar to the typical azimuthal
wavelength of the Parker instability, so the final scale of vertical variation
of the field likely comes about as the magnetic tension in the rising field
lines grows to match the buoyant rise of the field lines. The non-linear
transition from the initially stratified state to a state with magnetic arcs
anchored in the midplane leads to the formation of an approximately force-free
corona extending from $\sim$4 scale heights from the midplane. In the corona,
the magnetic tension and pressure terms are large (compared to their
difference) and oppositely directed, and current flows parallel to the magnetic
field lines. Increasing the vertical extent of the box from 12 to 24 scale
heights leads to a transport of some magnetic flux to regions with $|z|>6 H$,
at the cost of a 20-30\% decrease in the magnetic flux in the region $|z|<6
H$. The overall shape of the azimuthal magnetic field distribution is
nevertheless relatively unchanged.
\begin{figure}
  \includegraphics{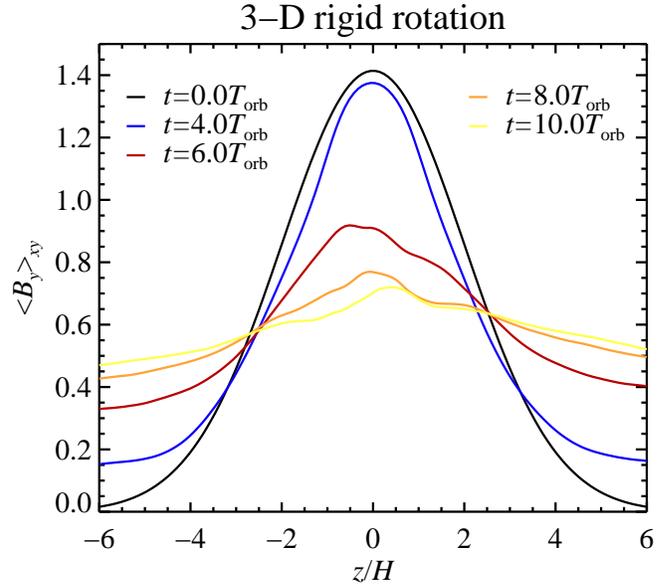}
  \caption{Mean azimuthal magnetic field versus height over the midplane for
    the Parker instability in 3-D rigid rotation. The azimuthal field quickly
    spreads evenly over the entire vertical extent of the box, in contrast to
    the 2-D case (shown in \Fig{f:Bymxy_z}) which makes a non-linear transition
    to a state that still has azimuthal flux differences.}
  \label{f:Bymxy_z_3D_noshear}
\end{figure}
\begin{figure*}
  \begin{center}
    \includegraphics[width=0.1\linewidth]{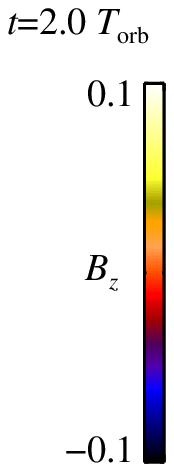}
    \includegraphics[width=0.35\linewidth]{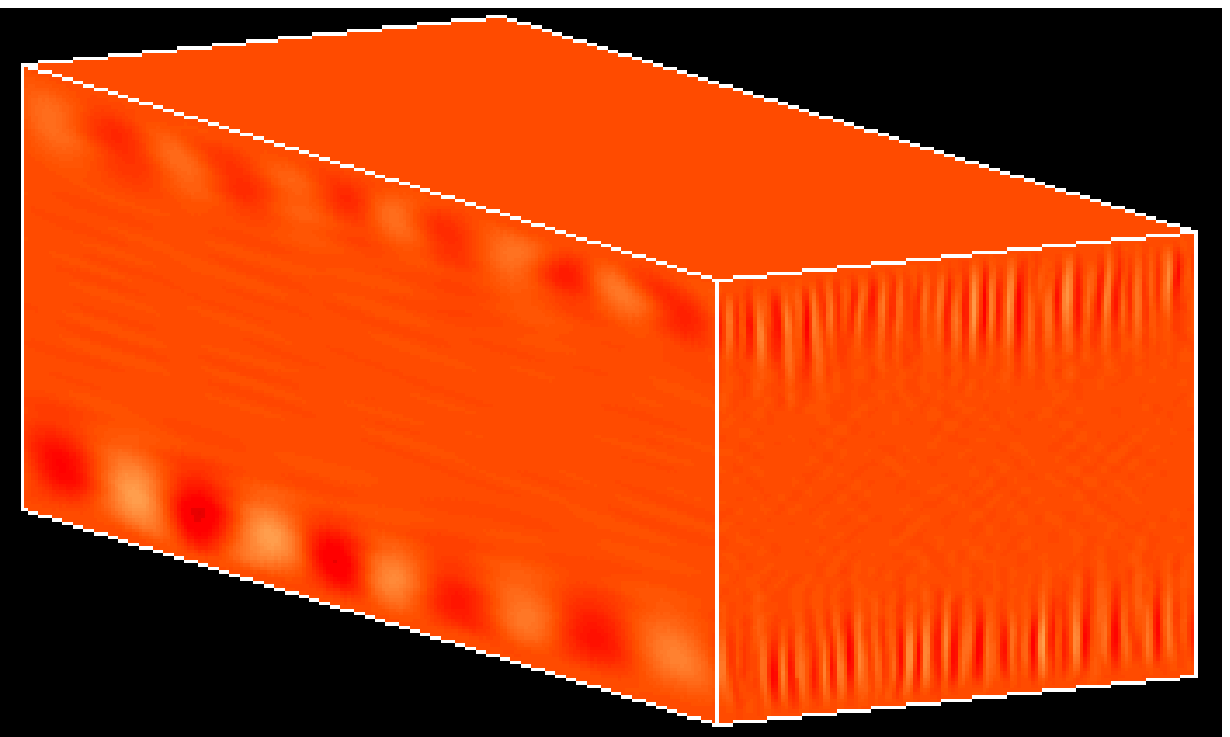}
    \includegraphics[width=0.1\linewidth]{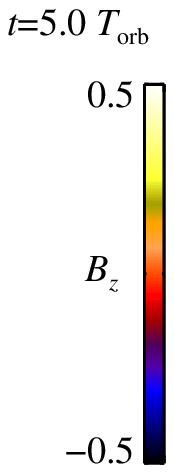}
    \includegraphics[width=0.35\linewidth]{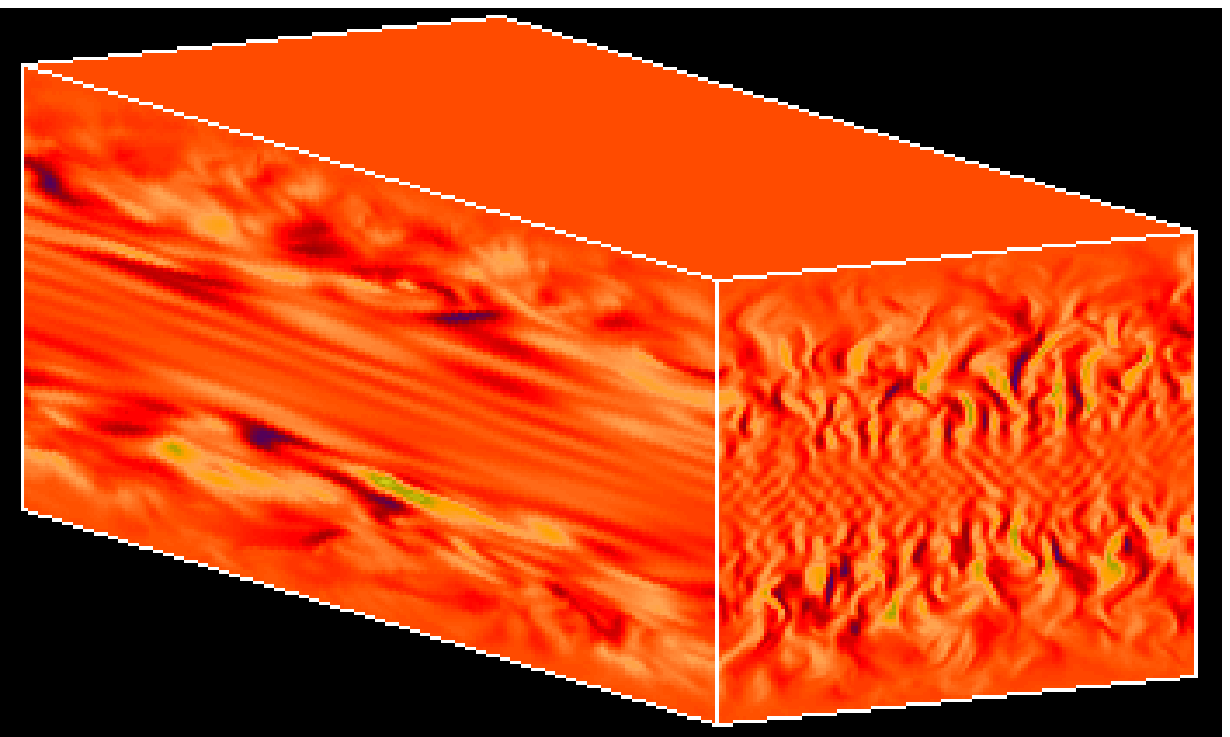}\\
    \includegraphics[width=0.1\linewidth]{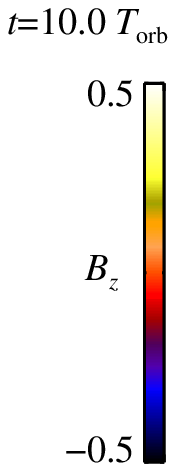}
    \includegraphics[width=0.35\linewidth]{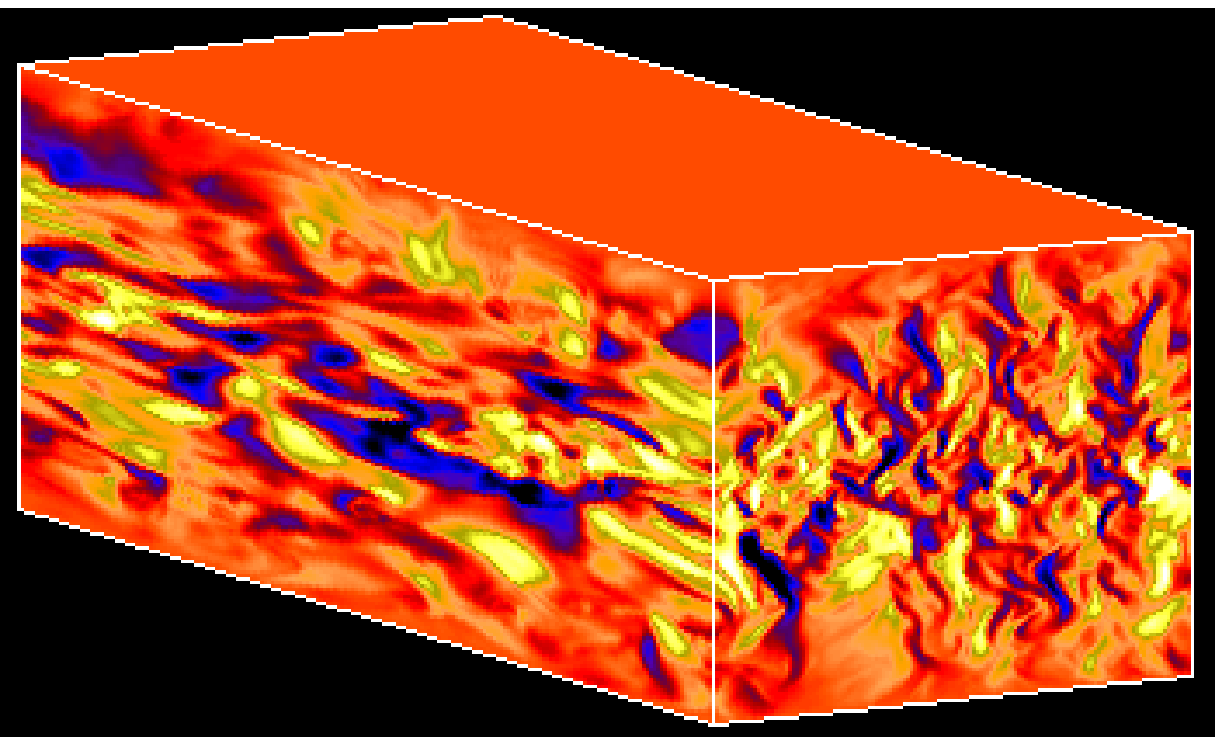}
    \includegraphics[width=0.1\linewidth]{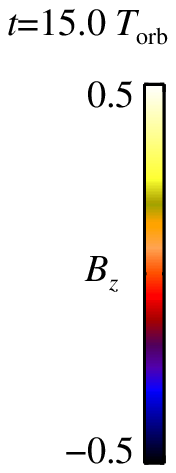}
    \includegraphics[width=0.35\linewidth]{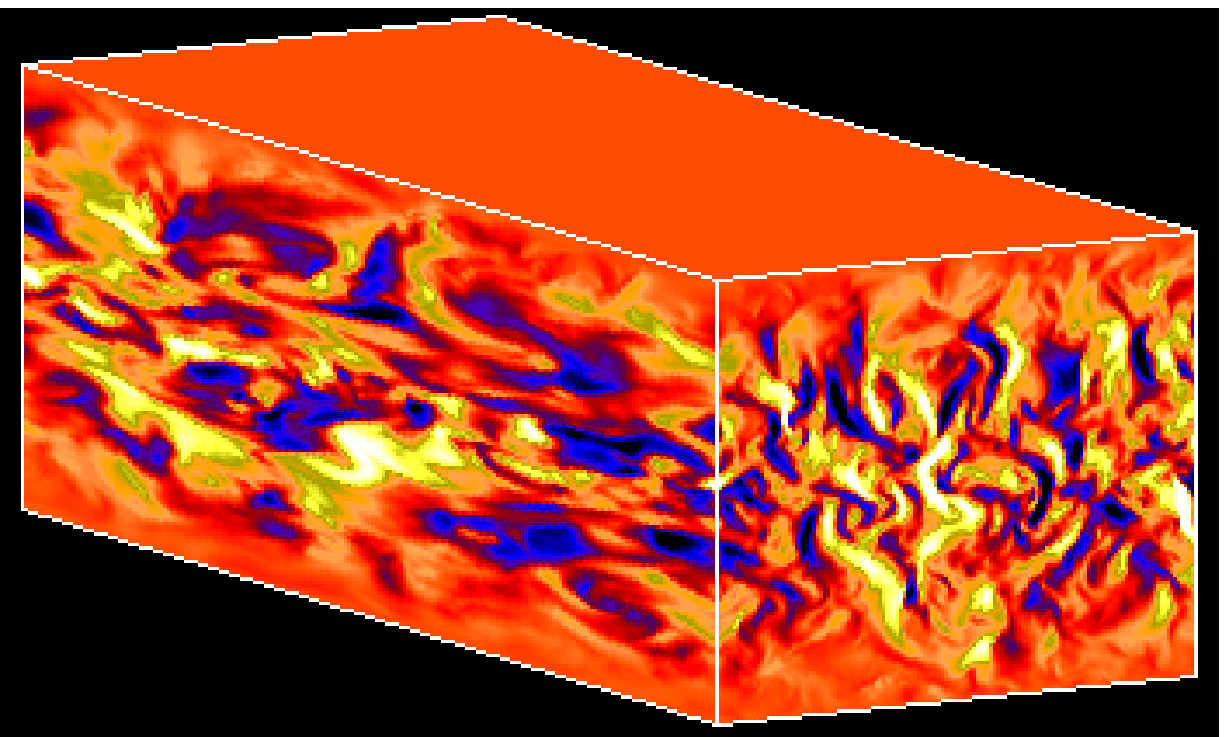}
  \end{center}
  \caption{The vertical magnetic field at the sides of the simulation box for
    run S3D\_512. The initial fastest growing wavelength of the Parker
    instability is around 5 scale heights in the azimuthal direction and 0.5
    scale heights in the radial direction. The magnetorotational instability
    eventually sets in as the vertical field grows in amplitude, and the box
    turns turbulent with a strongly fluctuating vertical field.}
  \label{f:parkerbox}
\end{figure*}

\subsection{Rigid rotation in 3-D}

Two-dimensional models of the Parker instability allow only the undular
mode to develop, but not the related interchange mode
\citep{Matsumoto+etal1993}, and hence the field has no possibility to escape
from the disc. In three-dimensional rigidly rotating disc simulations, the
field does escape. In \Fig{f:Bymxy_z_3D_noshear} we show the evolution of the
azimuthal field in a 3-D model including rotation (but still no shear). Now the
azimuthal field quickly redistributes itself evenly with height over the
midplane, leading to a state where vertical gravity is balanced only by
thermal pressure. Allowing the mixed undular/interchange instability to take
place leads to a genuinely turbulent state \citep{Kim+etal1998}, which mixes the
azimuthal field vertically and eventually spreads it uniformly throughout the
entire extent of the box. In the next section we will show, however, that
Keplerian shear completely suppresses the vertical spreading of the azimuthal
field, which remains concentrated towards the disc midplane due to a dynamo
coupling between $B_x$ and $B_y$.

The Parker instability in a shearing, rotating frame (which is the relevant
frame for most astrophysical purposes) must necessarily be more complicated
than in the inertial frame. The intrinsically non-axisymmetric instability is
deformed by the radial shear, while the whole coordinate frame can tap energy
from the gravitational potential through Maxwell and Reynolds stresses.
Already \cite{Shu1974} expanded the Parker instability to include the effect of
shear. The conclusion of the analytical stability analysis is that no amount
of shear can completely stabilise against the Parker instability, as the
instability can develop at such small radial scales that growth happens faster
than shear.

\section{The Parker instability in Keplerian shear}
\label{s:shear}

Having gained insight into the evolution of the Parker instability in 2-D and
3-D models with rotation but no shear, we now turn to the combined effect of
Parker and magnetorotational instabilities in a shearing sheet model of a
Keplerian accretion disc.

\subsection{Creation of vertical field}
\label{s:vfield}

In \Fig{f:parkerbox} we show the evolution of the vertical component of
the magnetic field at the sides of the box for run S3D\_512. The following
sequence of events is observed: first the (sheared) Parker instability grows
unaffected by the MRI, but as the vertical magnetic fields gets a significant
amplitude, the MRI sets in and leads to a turbulent state where both PI and MRI
determine the dynamics. The Parker instability initially sets in high up in
the atmosphere where the buoyancy is strongest, at the typical wavelength
$\lambda_y=5 H$ and $\lambda_x=0.5 H$. We actually observe that the radial
wavelength of the Parker instability decreases with increasing resolution as
the reduced dissipation allows growth at higher wave numbers. Eventually the
whole disc goes into a non-linear turbulent state with strongly fluctuating
vertical fields and twisted magnetic field loops extending far above the
midplane of the disc. The typical radial coherence scale in this state is much
longer than in the initial linear growth phase (compare $t=2.0 T_{\rm orb}$ to
$t=15.0 T_{\rm orb}$ in \Fig{f:parkerbox}).

\begin{figure}
  \begin{center}
    \includegraphics[width=0.9\linewidth]{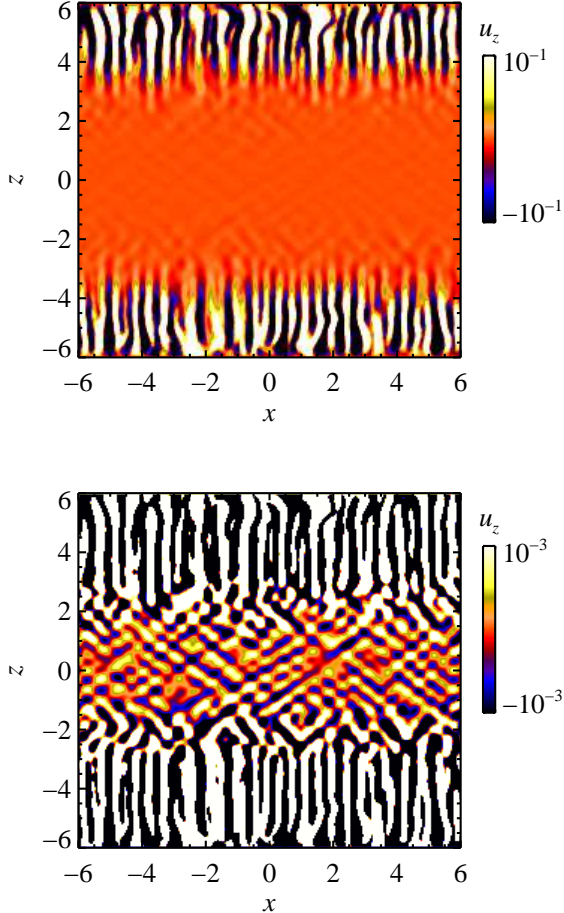}
  \end{center}
  \caption{The vertical velocity component at the $y=-L_y/2$ plane during
    the linear phase of run S3D\_512 (at $t=2.4 T_{\rm orb}$). The top panel
    shows the high radial wavenumber structures typical for the 3-D Parker
    instability. However, changing the contrast by two orders of magnitude
    reveals structures of high {\it vertical} wavenumber in the midplane of
    the disc, likely a result of magnetorotational instability in the azimuthal
    field.}
  \label{f:uz_xz}
\end{figure}

The initial azimuthal field is also unstable to a (transient)
non-axisymmetric magnetorotational ``instability''
\citep{BalbusHawley1992,FoglizzoTagger1995,TerquemPapaloizou1996,Keppens+etal2002}.
Magnetorotational instability in the azimuthal field has highest growth rate
for modes with a small vertical wavelength\footnote{It was shown by
\cite{FoglizzoTagger1994,FoglizzoTagger1995} that in the $\beta=1$ case the
azimuthal field MRI should have a slightly higher growth rate than the Parker
instability. However, their linear stability analysis assumed a constant
gravity field, rather than a linear field which is the more relevant for
Keplerian discs \citep{Kim+etal1997}. Hence the growth rate of the Parker
instability several scale heights from the midplane may be underestimated in
\cite{FoglizzoTagger1994}.}. The absence of high $k_z$ modes in the first
panel of \Fig{f:parkerbox} indicates that non-axisymmetric MRI is not the
dominant driver of the linear dynamics of the system. However, in \Fig{f:uz_xz}
we show the vertical velocity field at the $y=-L_y/2$ plane for run S3D\_512
during the linear growth phase. The
top panel exhibits the usual signs of Parker instability high up in the
atmosphere, but changing the contrast by two orders of magnitude (bottom panel)
reveals modes of very short radial and vertical wavelength in the midplane of
the disc, representing trailing waves that have already been amplified by
magnetorotational instability in the azimuthal field. Transient amplification
of swinging waves may also be important in determining the non-linear evolution
of the system at later times \citep[e.g.][]{Rincon+etal2007,LesurOgilvie2008}.

\begin{figure}
  \begin{center}
    \includegraphics[width=\linewidth]{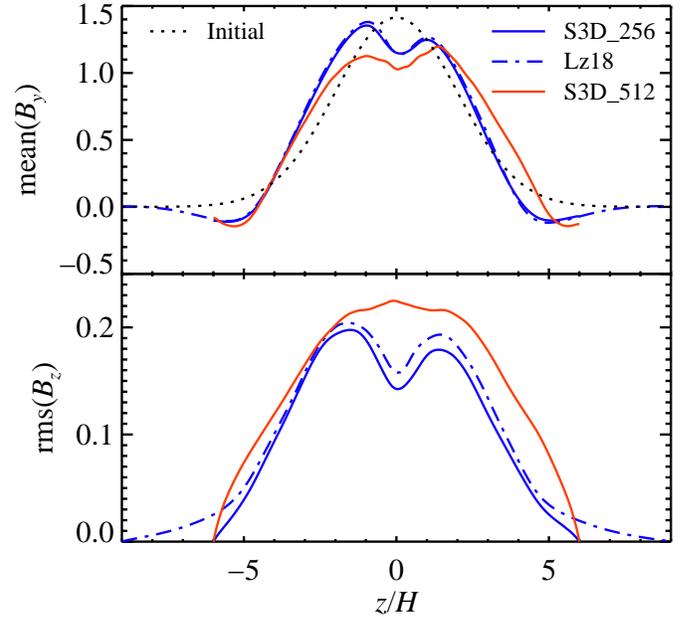}
  \end{center}
  \caption{Confinement of the azimuthal field in shearing box simulations.
    The mean azimuthal field and the root-mean-square of the vertical magnetic
    field are plotted as a function of height over the midplane. The Parker
    instability creates strong vertical fields from the initial azimuthal flux
    (dotted line in top plot). Vertical loss of azimuthal flux is nevertheless
    completely suppressed outside of a field reversal occurring at
    approximately four scale heights from the midplane.
    Doubling the resolution (S3D\_512) leads
    to an increased vertical field strength and a decreased azimuthal field in
    the midplane, due to a significant increase in turbulent motions at higher
    resolution (see \Tab{t:results}).}
  \label{f:bymean_bzrms_z}
\end{figure}

Although production of vertical field is not a main ingredient of
azimuthal MRI, any unstable mode with a non-axisymmetric vertical velocity
component can create significant vertical field by stretching the large scale
azimuthal field (although highly variable in the vertical direction). The
production of large scale vertical field by the Parker instability, on the
other hand, is demonstrated in Figure \ref{f:bymean_bzrms_z}. Here we plot the
mean azimuthal field and the root-mean-square of the vertical field as a
function of height over the midplane for the mixed PI/MRI shearing sheet
simulations. Significant vertical magnetic fields of magnitude $B_z$$\sim$$0.2$
(with associated pressure $\beta_z$$\sim$$50$) arise due to magnetic buoyancy.
There is a reversal of the azimuthal magnetic field at around four scale
heights from the midplane. The correspondence between run S3D\_256 and run
S3D\_256\_Lz18 is extremely close, so the sign reversal is not due to the
presence of the vertical boundary. Increasing resolution leads to some
increase in vertical field strength in the midplane, but the field reversal
still takes place at the same height over the midplane. Although the
non-linear state of the combined PI and MRI is time-dependent and fluctuating,
there is a clear saturation and vertical confinement of the mean azimuthal
magnetic field (we return to the issue of field confinement in
\S\ref{s:confinement}).

\subsection{Stresses}

\begin{figure}
  \begin{center}
    \includegraphics[width=\linewidth]{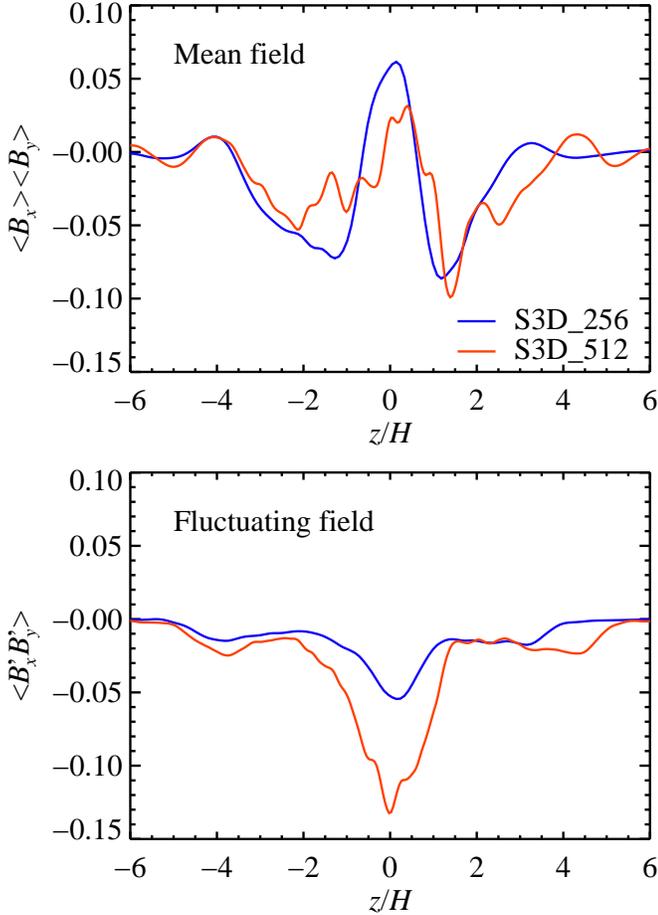}
  \end{center}
  \caption{The Maxwell stress $\langle B_x B_y \rangle$ as a function of height
    over the midplane. The top panel shows the stress from the mean field
    $\langle B_x \rangle \langle B_y \rangle$, while the bottom panel shows the
    contribution from the fluctuating field
    $\vc{B}'=\vc{B}-\langle\vc{B}\rangle$. The magnetorotational
    instability feeds off both the azimuthal fields and the large scale vertical
    fields created by the Parker instability, to create significant stresses,
    up to $\langle B_x B_y \rangle\approx-0.1$ in the midplane of the high
    resolution run S3D\_512.}
  \label{f:bxbym_z}
\end{figure}

The magnetorotational instability in turn feeds off both the vertical
field lines created by the Parker instability and the azimuthal field component
still present in the non-linear state. The resulting Maxwell stress $\langle
B_x B_y \rangle$, averaged over time and radial and azimuthal directions, is
shown as a function of height over the midplane in Figure \ref{f:bxbym_z}. We
have divided the stress into contributions from the mean field, $\langle
\langle B_x \rangle_{xy} \langle B_y \rangle_{xy} \rangle_t$, and fluctuating
field, $\langle (B_x-\langle B_x \rangle_{xy}) (B_y-\langle B_y \rangle_{xy})
\rangle_{xyt}$. The occurrence of a large scale $B_x$ is discussed in detail in
\S\ref{s:confinement}. The radial field couples with the large scale azimuthal
field to yield a large scale structure in the Maxwell stress as well,
relatively well converged when increasing the resolution.

The fluctuating magnetic field is associated with significant stresses around
the midplane, with a Maxwell stress in the range $0.03\ldots0.1$ in the
regions within a couple of scale heights from the midplane. The stress from
this fluctuating field increases significantly with increasing resolution, as
the decreasing dissipation on small radial scales allows the Parker instability
to create stronger vertical fields, while the decreased numerical and
artificial dissipation lets the both vertical field and azimuthal field MRI
develop faster.

\begin{figure}
  \begin{center}
    \includegraphics[width=\linewidth]{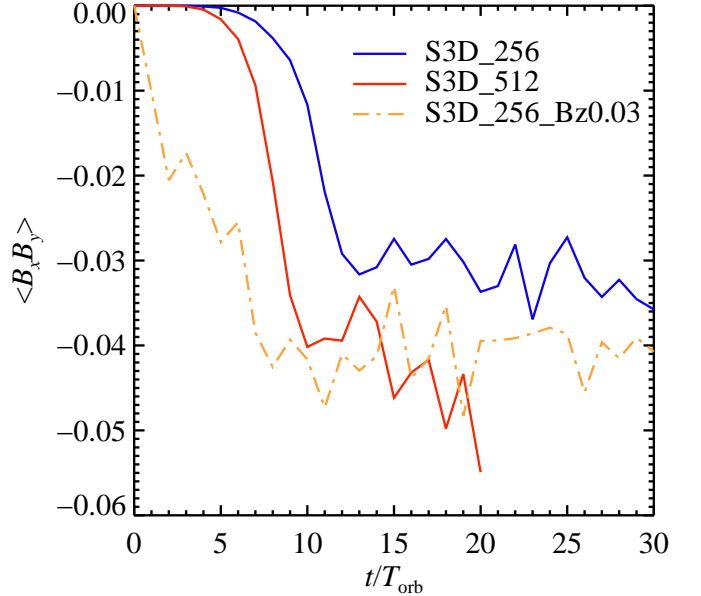}
  \end{center}
  \caption{The mean Maxwell stress as a function of time. The high resolution
    run S3D\_512 has faster growth of the stress and higher saturation level.
    Imposing the gas to a moderate net vertical field (dash-dotted line) lets
    the vertical field MRI develop from the beginning, but eventually the
    stresses saturate at an only slightly higher level (in its absolute value)
    than in the zero net vertical flux case.}
  \label{f:bxbym_t}
\end{figure}

In \Fig{f:bxbym_t} we show the measured mean Maxwell stress $\langle B_x B_y
\rangle$ as a function of time. The run S256\_3D\_Bz0.03 has a moderate
vertical field ($B_z=0.03$) imposed through the box\footnote{We exclude the
vertical field from the top and bottom six grid zones, to avoid conflict with
the zero-vertical-field boundary condition.}. With this set up the
magnetorotational instability sets in before the Parker instability does,
creating significant stresses already after a few orbits. Eventually however
the Parker instability develops as usual, and the stresses reach saturation at
a level that is only slightly higher (in absolute value) than for the zero net
vertical flux run. Thus it seems that it is not very important whether the
magnetorotational instability develops before the Parker instability or vice
versa. This situation may nevertheless change when going to either weaker
azimuthal fields or stronger vertical fields, in which case the turbulent
diffusion created by the magnetorotational instability may reduce the midplane
azimuthal flux quicker than the Parker instability can grow \cite[but
see][where the Parker instability arises in simulations with a net azimuthal
field that is much weaker than ours]{Blaes+etal2007}. In the case of a strong,
imposed vertical field, there is however already an in inexhaustible source of
accretion stresses present in the disc without the need to invoke an additional
mechanism based on azimuthal fields and Parker instability.

In \Tab{t:results} we show the measured statistical properties of the
non-linear state of the combined Parker and magnetorotational instabilities. We
divide into the midplane regions with $|z|<2$ (containing 85\% of the gas
mass) and atmosphere regions with $|z|>2$ (containing 15\% of the gas mass).
The midplane regions are characterised by strong magnetic fields and high
accretion through magnetic stresses, whereas the atmosphere has stronger
velocity fields and higher Reynolds stress, but weaker magnetic fields and
Maxwell stresses. Decreasing the initial magnetic pressure by a factor of three
(run S3D\_256\_b3) leads to an expected decrease in both magnetic energies and
stresses.

The $z$-dependent Reynolds and Maxwell stresses can be translated into an
average turbulent viscosity \citep[following][]{Brandenburg+etal1995},
\begin{eqnarray}
  \langle \rho u_x u_y \rangle &=& \frac{3}{2} \varOmega \nu_{\rm t}^{\rm
  (kin)} \langle \rho \rangle \, , \label{eq:nutkin} \\
  -\frac{1}{\mu_0} \langle B_x B_y \rangle &=& \frac{3}{2} \varOmega \nu_{\rm
  t}^{\rm (mag)} \langle \rho \rangle \, . \label{eq:nutmag}
\end{eqnarray}
Here $\nu_{\rm t}^{\rm (kin)}$ and $\nu_{\rm t}^{\rm (mag)}$ are the turbulent
viscosities due to the velocity field and the magnetic field, respectively. We
can further normalise the turbulent viscosities by the sound speed $c_{\rm s}$
and gas scale height $H$ \citep{ShakuraSunyaev1973,Pringle1981},
\begin{eqnarray}
  \nu_{\rm t}^{\rm (kin)} &=& \alpha_{\rm t}^{\rm (kin)} c_{\rm s} H \, ,
  \label{eq:alphakin} \\
  \nu_{\rm t}^{\rm (mag)} &=& \alpha_{\rm t}^{\rm (mag)} c_{\rm s} H \, .
  \label{eq:alphamag} 
\end{eqnarray}
For the gas scale height we use the expression
$H=H_\beta=\sqrt{1+\beta^{-1}}c_{\rm s}/\varOmega$, with $\beta\equiv P/P_{\rm
mag}$ given by the initial value. The resulting $\alpha$-values are given in
\Tab{t:viscosity}. The large scale vertical fields arising from the Parker
instability, together with a non-axisymmetric instability in the
azimuthal field, give rise to $\alpha_{\rm t}\approx0.05\ldots0.1$, resulting
in very high accretion rates through $\dot{M}=3\pi\nu_{\rm t}\varSigma_{\rm
gas}$ \citep[][]{Pringle1981}.

\cite{Machida+etal2000} observed similarly high $\alpha$-values in their
global simulation of a strongly magnetised accretion disc, and they attributed
the Maxwell stresses to a
non-axisymmetric magnetorotational instability of the azimuthal field.
\cite{Kim+etal2002} considered the competition between Parker and gravitational
instabilities in the context of a galactic potential. Simulations including
rotation and shear nevertheless did not show any sign of the magnetorotational
instability and the authors note that the magnetorotational instability in the
azimuthal field grows too slowly to show up in the simulation time-scale of 4-5
orbits. In this work we find that the onset of turbulence fits a two layer
scenario where the atmosphere is dominated by Parker instability and MRI in the
ensuing vertical fields, whereas azimuthal field MRI drives the linear growth
within a couple of scale heights of the disc midplane.
\begin{table*}[!t]
  \caption{Statistical flow properties in the saturated, turbulent state.}
  \begin{center}
    $|z|<2 H$ (85\% of the mass)\\
    {\tiny
    \begin{tabular}{lccccccccc}
      \hline
        Run & ${\rm rms}(u_x)$ &
              ${\rm rms}(u_y)$ &
              ${\rm rms}(u_z)$ &
              ${\rm mean}(u_x u_y)$ &
              ${\rm rms}(B_x)$ &
              ${\rm rms}(B_y)$ &
              ${\rm rms}(B_z)$ &
              ${\rm mean}(B_x B_y)$ & \\
      \hline
        S3D\_256 &
        $ 0.346 \pm 0.020$ & $ 0.261 \pm 0.020$ & $ 0.344 \pm 0.016$ & 
        $ 0.017 \pm 0.005$ & $ 0.189 \pm 0.009$ & $ 1.227 \pm 0.028$ & 
        $ 0.172 \pm 0.005$ & $-0.055 \pm 0.006$ \\
        S3D\_512 &
        $ 0.366 \pm 0.017$ & $ 0.331 \pm 0.009$ & $ 0.338 \pm 0.025$ & 
        $ 0.031 \pm 0.003$ & $ 0.287 \pm 0.015$ & $ 1.143 \pm 0.027$ & 
        $ 0.230 \pm 0.012$ & $-0.087 \pm 0.016$ \\
        S3D\_256\_b3 &
        $ 0.341 \pm 0.017$ & $ 0.232 \pm 0.009$ & $ 0.359 \pm 0.021$ & 
        $ 0.011 \pm 0.001$ & $ 0.142 \pm 0.009$ & $ 0.801 \pm 0.016$ & 
        $ 0.128 \pm 0.006$ & $-0.040 \pm 0.005$ \\
        S3D\_256\_Lz18 &
        $ 0.400 \pm 0.028$ & $ 0.314 \pm 0.013$ & $ 0.370 \pm 0.030$ & 
        $ 0.025 \pm 0.003$ & $ 0.208 \pm 0.011$ & $ 1.240 \pm 0.043$ & 
        $ 0.187 \pm 0.013$ & $-0.058 \pm 0.010$ \\
        S3D\_256\_Bz0.03 &
        $ 0.395 \pm 0.024$ & $ 0.292 \pm 0.013$ & $ 0.408 \pm 0.012$ & 
        $ 0.021 \pm 0.002$ & $ 0.216 \pm 0.011$ & $ 1.347 \pm 0.018$ & 
        $ 0.207 \pm 0.003$ & $-0.084 \pm 0.012$ \\
      \hline
    \end{tabular}}
    \\\vspace{0.5cm}
    $|z|>2 H$ (15\% of the mass)\\
    {\tiny
    \begin{tabular}{lccccccccc}
      \hline
        Run & ${\rm rms}(u_x)$ &
              ${\rm rms}(u_y)$ &
              ${\rm rms}(u_z)$ &
              ${\rm mean}(u_x u_y)$ &
              ${\rm rms}(B_x)$ &
              ${\rm rms}(B_y)$ &
              ${\rm rms}(B_z)$ &
              ${\rm mean}(B_x B_y)$ & \\
      \hline
        S3D\_256 &
        $ 1.108 \pm 0.084$ & $ 0.696 \pm 0.042$ & $ 0.635 \pm 0.016$ & 
        $ 0.059 \pm 0.066$ & $ 0.116 \pm 0.007$ & $ 0.486 \pm 0.015$ & 
        $ 0.100 \pm 0.005$ & $-0.016 \pm 0.004$ \\
        S3D\_512 &
        $ 1.072 \pm 0.086$ & $ 0.793 \pm 0.036$ & $ 0.665 \pm 0.035$ & 
        $ 0.096 \pm 0.026$ & $ 0.163 \pm 0.004$ & $ 0.547 \pm 0.016$ & 
        $ 0.138 \pm 0.006$ & $-0.025 \pm 0.004$ \\
        S3D\_256\_b3 &
        $ 1.159 \pm 0.125$ & $ 0.634 \pm 0.075$ & $ 0.597 \pm 0.067$ & 
        $ 0.022 \pm 0.065$ & $ 0.059 \pm 0.005$ & $ 0.194 \pm 0.010$ & 
        $ 0.044 \pm 0.003$ & $-0.004 \pm 0.001$ \\
        S3D\_256\_Lz18   &
        $ 1.439 \pm 0.092$ & $ 0.956 \pm 0.029$ & $ 0.897 \pm 0.077$ & 
        $ 0.130 \pm 0.054$ & $ 0.100 \pm 0.005$ & $ 0.390 \pm 0.017$ & 
        $ 0.085 \pm 0.006$ & $-0.012 \pm 0.002$ \\
        S3D\_256\_Bz0.03 &
        $ 1.403 \pm 0.101$ & $ 0.832 \pm 0.058$ & $ 0.735 \pm 0.033$ & 
        $ 0.192 \pm 0.099$ & $ 0.135 \pm 0.009$ & $ 0.448 \pm 0.027$ & 
        $ 0.107 \pm 0.006$ & $-0.016 \pm 0.004$ \\
      \hline
    \end{tabular}}
  \end{center}
  The two tables indicate the root-mean-square (and its temporal fluctuation
  interval) of the velocity field and the magnetic field, and the mean product
  of the radial and azimuthal components of velocity field and magnetic field.
  The top table shows the values within two scale heights of the midplane,
  while the bottom table shows the values measured above two scale heights.
  \label{t:results}
\end{table*}

\subsection{Energy spectra}

\begin{figure}
  \begin{center}
    \includegraphics[width=\linewidth]{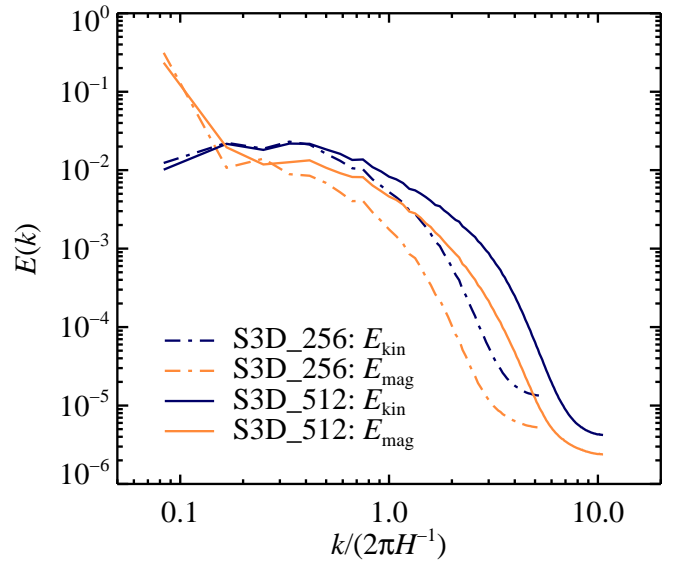}
  \end{center}
  \caption{Kinetic and magnetic energy spectrum for runs S3D\_256 and S3D\_512.
    The power $P(\vc{k})$ has been averaged over snapshots between 15 and 20
    orbits and summed over shells of constant wave number $k=|\vc{k}|$. Kinetic
    energy dominates over magnetic energy at most scales, except for the few
    largest scale of the box.}
  \label{f:energy_spectrum}
\end{figure}

The kinetic and magnetic energy spectra are shown in \Fig{f:energy_spectrum}
for runs S3D\_256 and S3D\_512. We have taken the power at scale $\vc{k}$,
$P(\vc{k})$, and summed over shells of constant wave number $k=|\vc{k}|$,
excluding the anisotropic scale $k=2\pi/(24 H)$ that is only present in the
$y$-direction. \Fig{f:energy_spectrum} shows that kinetic energy dominates over
magnetic energy by approximately a factor of two at most scales, except for the
1-2 largest scales which are dominated by the magnetic field.

\begin{table*}[!t]
  \begin{center}
    \begin{tabular}{lcccccc}
      \hline
        Run & $\nu_{\rm t}^{\rm (mag)}$ &
              $\nu_{\rm t}^{\rm (kin)}$ &
              $\nu_{\rm t}^{\rm (tot)}$ &
              $\alpha_{\rm t}^{\rm (mag)}$ &
              $\alpha_{\rm t}^{\rm (kin)}$ &
              $\alpha_{\rm t}^{\rm (tot)}$ \\
      \hline
            S3D\_256         & $0.075$ & $0.020$ & $0.095$ &
                               $0.053$ & $0.014$ & $0.067$ \\
            S3D\_512         & $0.107$ & $0.033$ & $0.141$ &
                               $0.076$ & $0.023$ & $0.100$ \\
            S3D\_256\_b3     & $0.042$ & $0.010$ & $0.052$ &
                               $0.030$ & $0.007$ & $0.037$ \\
            S3D\_256\_Lz18   & $0.078$ & $0.017$ & $0.095$ &
                               $0.055$ & $0.012$ & $0.067$ \\
            S3D\_256\_Bz0.03 & $0.096$ & $0.023$ & $0.120$ &
                               $0.068$ & $0.017$ & $0.085$ \\
      \hline
    \end{tabular}
  \end{center}
  \caption{Turbulent viscosity coefficients and $\alpha$-values, based on
    \Eqss{eq:nutkin}{eq:alphamag}.}
  \label{t:viscosity}
\end{table*}

\section{Field confinement}
\label{s:confinement}

\begin{figure}
  \begin{center}
    \includegraphics[width=\linewidth]{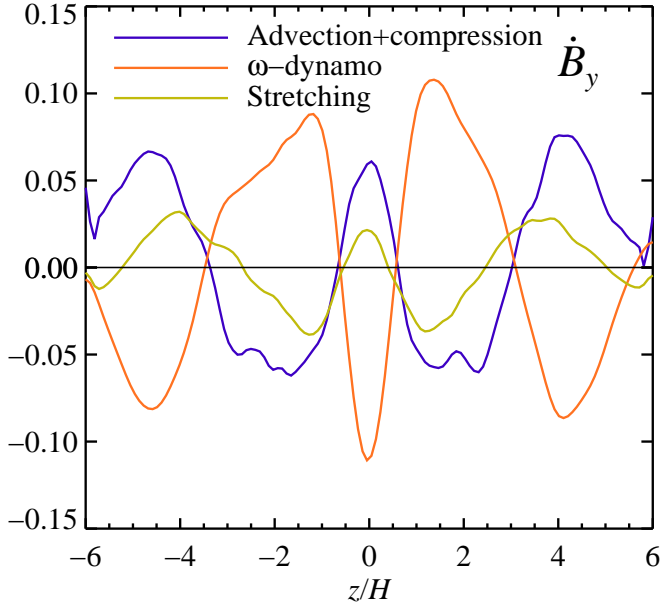}
  \end{center}
  \caption{Different terms in the azimuthal component of the induction equation
    averaged over $x$ and $y$ and over evenly spaced snapshots between
    $t=10T_{\rm orb}$ and $t=30T_{\rm orb}$, for run S3D\_256. The azimuthal
    magnetic field is in equilibrium between turbulent transport, representing
    compression and advection due to both the mean and the fluctuating velocity
    field (turbulent resistivity), and the Keplerian shear $\omega$-dynamo that
    creates $B_y$ out of $B_x$.}
  \label{f:vertical_equilibrium_By}
\end{figure}
\begin{figure}
  \begin{center}
    \includegraphics[width=\linewidth]{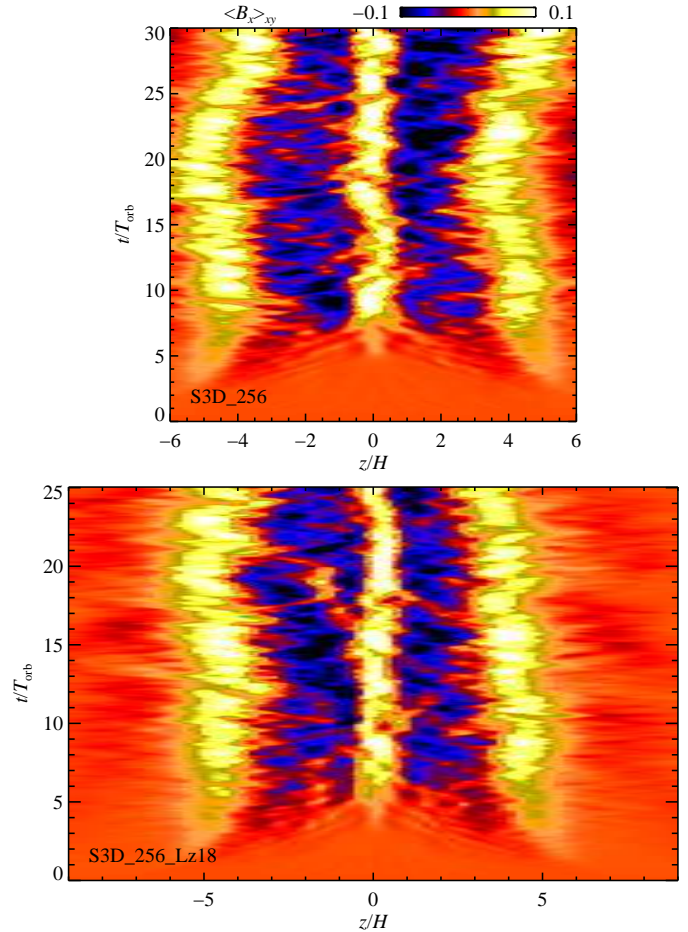}
  \end{center}
  \caption{The radial component of the magnetic field, $B_x$, averaged over the
    radial and azimuthal directions, for runs with $L_z=12 H$ (top panel) and
    $L_z=18H$ (bottom panel). The vertical structure quickly develops a peak in
    the midplane and one additional peak on each side of the midplane, and
    this structure stays statistically unchanged for the duration of the
    simulation. This radial field drives an $\omega$-dynamo that balances the
    turbulent resistivity acting on the mean azimuthal field (see
    \Fig{f:vertical_equilibrium_By}).}
  \label{f:Bx_z_t}
\end{figure}
\begin{figure}
  \begin{center}
    \includegraphics[width=\linewidth]{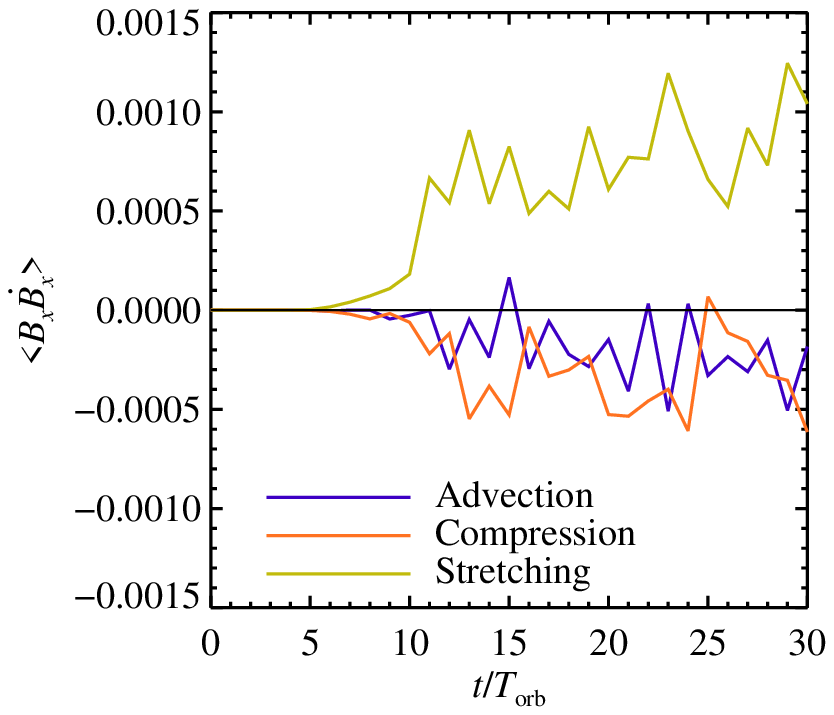}
  \end{center}
  \caption{The power contribution to the radially and azimuthally averaged
    radial field, as a function of time measured in orbits. Power is provided
    almost exclusively by the magnetic stretching term $\vc{B}\cdot\nab u_x$,
    while advection and compression are both sink terms.}
  \label{f:vertical_equilibrium_power_Bx}
\end{figure}
An important issue related to our proposed path to accretion is whether the
original azimuthal flux can stay confined in the disc or whether it will escape
to infinity by the action of turbulent resistivity (i.e.\ magnetic buoyancy).
The combination of Parker and interchange instabilities in the non-shearing
frame will eventually redistribute the azimuthal magnetic field evenly over the
entire box height (see \Fig{f:Bymxy_z_3D_noshear}). We emphasise that we do not
see the same behaviour in the shearing sheet. The magnetic field and
velocity field stay statistically constant for the entire duration of the
simulations, with no sign of decay or gradual loss of azimuthal flux
(\Fig{f:bymean_bzrms_z}).

\subsection{Vertical structure of the field}
\label{s:bz}

In \Fig{f:vertical_equilibrium_By} we dissect the equilibrium state of $B_y$ by
averaging different terms from the induction equation over $x$ and $y$ and over
evenly spaced snapshots between $t=10T_{\rm orb}$ and $t=30T_{\rm orb}$. There
is an almost perfect equilibrium between turbulent transport of magnetic fields
[$\dot{B}_y=-\nab\cdot(\vc{u} B_y)$, representing compression and advection due
to both the mean part of the velocity field and the fluctuating part
(``turbulent resistivity'')] and azimuthal field created by the shear from the
radial field [$\dot{B}_y=-(3/2)\varOmega B_x$]. This $\omega$-dynamo allows for
a closure of the entire process by which strong accretion occurs from initially
purely azimuthal fields. The magnetic stretching term $(\vc{B}\cdot\nab)u_y$
(which does not include stretching by the Keplerian shear because we measure
velocities relative to the main shear) is generally of the same sign as the
turbulent transport term and thus acts oppositely of the shear term.

The vertical and temporal dependence of the radial magnetic field $B_x$ is
shown in \Fig{f:Bx_z_t} as a function of height over the midplane $z$ and time
$t$ measured in orbits. After around 5 orbits a clear structure evolves out of
the flow, with a peak of $B_x$ in the midplane of the disc, followed by a
valley and an additional peak at each side of the midplane. This structure
stays statistically unchanged for the entire duration of the run. A similar
steady state structure of the radial field was observed by
\cite{Hanasz+etal2002} in rigid rotation simulations of the galactic Parker
instability. In the simulations of \cite{Hanasz+etal2002} $B_x$ would be either
positive or negative in the midplane, depending on the radial wave number of
the initial perturbation. In the shearing sheet the radial wave number of the
PI is forced by the shear to be quite high, and our observation of a positive
radial field in the midplane agrees with the high radial wave number
simulations of \cite{Hanasz+etal2002}. Because we include shear in our
simulations we additionally see the constant creation of a $z$-dependent $B_y$,
by the Keplerian shear, which balances out the turbulent transport and prevents
the azimuthal field from spreading evenly over the box height.

In \Fig{f:vertical_equilibrium_power_Bx} we show the power contribution to the
large scale $B_x$ as a function of time. We have first averaged $B_x$ and the
advection, compression and stretching terms of the radial component of the
induction equation over the $x$- and $y$-directions. The power contribution of
each term is extracted by multiplying the time derivative $\dot{B}_x$ with
$B_x$ and averaging over $z$. Power is provided almost exclusively by the
magnetic stretching term $\dot{B}_x=\vc{B}\cdot\nab u_x$, whereas both
advection and compression extract energy from $B_x$ at all times (except for a
few peaks to around zero power). The magnetic stretching term provides a direct
coupling between $B_x$ and $B_y$, indicating that $B_x$ is created from $B_y$
in a dynamo process.
\begin{figure*}
  \begin{center}
    \includegraphics[width=\linewidth]{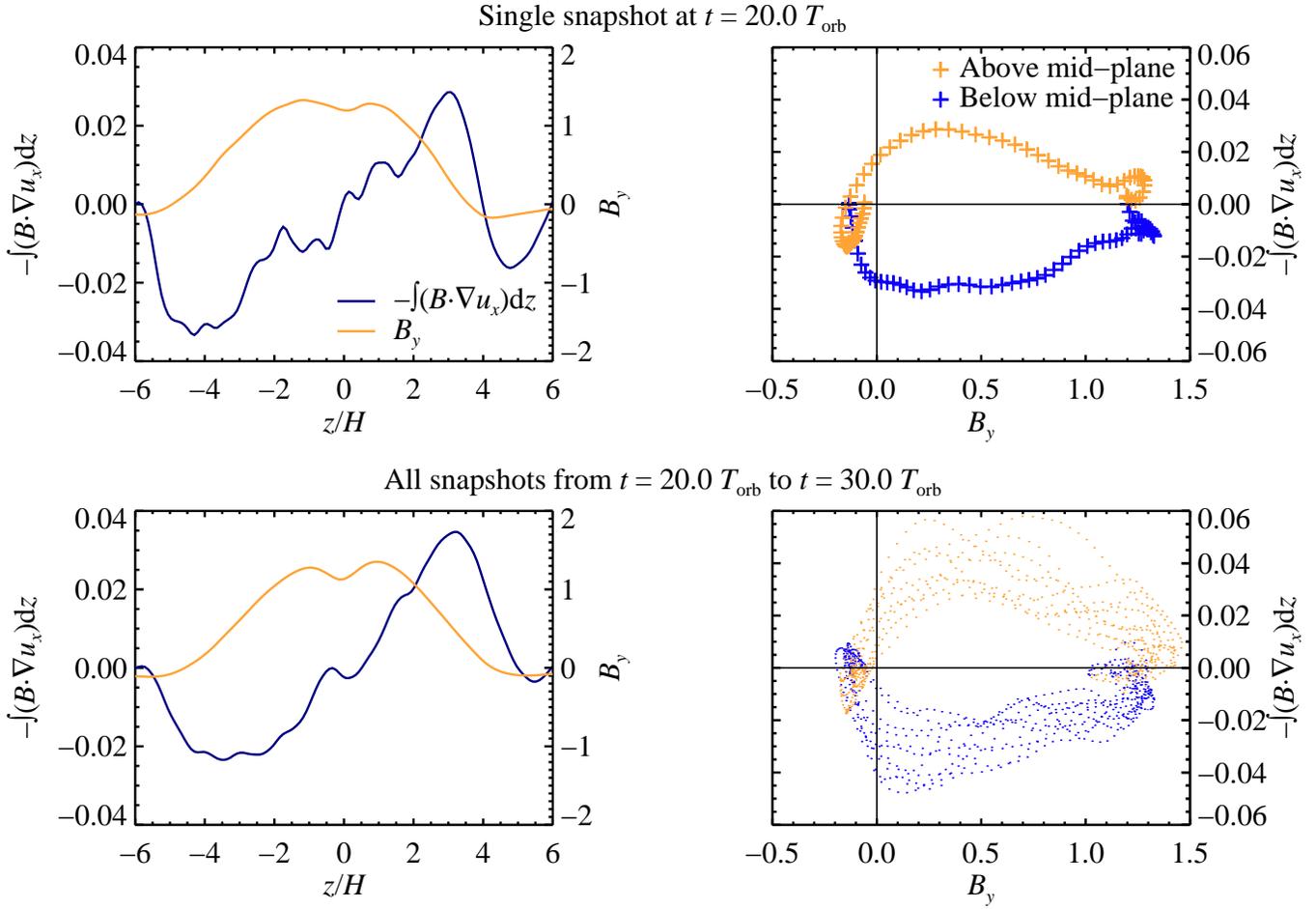}
  \end{center}
  \caption{Derivation of the dynamo $\alpha_{\rm dyn}$ from a single snapshot
    (top panels) and from the average of several snapshots (bottom panels). The
    (negative) integral of the magnetic stretching term $-\int
    \overline{\vc{B}\cdot\nab u_x}\de z$ is shown together with the azimuthal
    magnetic field $\overline{B}_y$, both as functions of height over the
    midplane, in the left panels. The electromotoric force contribution from
    the stretching term is of opposite sign on each side of the midplane. The
    right panels show the correlation between $\overline{B}_y$ and $-\int
    \overline{\vc{B}\cdot\nab u_x}\de z$, with orange (grey) denoting points
    above the midplane and blue (dark) denoting points below. There is a very
    clear correlation (anticorrelation) above (below) the midplane, indicating
    a dynamo $\alpha$ that is positive above the midplane and negative below,
    and of order $\alpha_{\rm dyn}\approx0.023c_{\rm s}$.}
  \label{f:alphadyn}
\end{figure*}

\subsection{Dynamo}

The creation of a systematic radial field component can be understood from gas
that plunges down field lines that have been bent by the Parker
instability \citep{Parker1992,HanaszLesch1993,Hanasz+etal2004}. As gas slides
azimuthally along the field lines, the Coriolis force causes a counter
clockwise rotation around dense clumps and clockwise rotation around underdense
regions, twisting magnetic field lines in such a way that the perturbed
electromotoric force points parallel to the (unperturbed) azimuthal field
line. Radial field is subsequently created from the stretching of the
perturbed field lines, in much the same way as imagined for the canonical
$\alpha$-mechanism \citep{Parker1955,Moffatt1978}.

To explore this scenario in more quantitative terms we follow
\cite{Moffatt1978} and expand the velocity and magnetic fields in constant and
fluctuating parts, $\vc{u}=\overline{\vc{U}}+\vc{u}'$,
$\vc{B}=\overline{\vc{B}}+\vc{b}'$, leading to the following evolution equation
for the mean magnetic field
\begin{equation}
  \dot{\overline{\vc{B}}} =
  \nab\times(\overline{\vc{U}}\times\overline{\vc{B}} + \alpha_{\rm
  dyn}\overline{\vc{B}} - \eta_{\rm t} \overline{\vc{J}}) \, .
  \label{eq:meanfield}
\end{equation}
Here overlines denote azimuthal and radial (and possible temporal) averaging,
while $\alpha_{\rm dyn}$ is a parameter that describes the proportionality
between mean field $\overline{\vc{B}}$ and fluctuating electromotoric force
$\overline{\vc{u}'\times\vc{b}'}$. The parameter $\eta_{\rm t}$ represents
turbulent resistivity, proportional to the current density
$\overline{\vc{J}}=\nab\times\overline{\vc{B}}$ of the mean field. Writing out
the radial component of \Eq{eq:meanfield} we get
\begin{equation}
  \dot{\overline{B}}_x = -\frac{\dpa}{\dpa z}(\overline{U}_z\overline{B}_x)-
  \frac{\dpa}{\dpa z}(\alpha_{\rm dyn} \overline{B}_y)
  + \eta_{\rm t} \frac{\dpa^2\overline{B}_y}{\dpa z^2}\, .
\end{equation}
The first term is due to advection of the large scale radial field by the large
scale vertical velocity component. \Fig{f:vertical_equilibrium_power_Bx}
showed that the source of magnetic energy in the $x$-component is the magnetic
stretching term. We may thus identify the positive contribution to
$\dot{\overline{B}}_x$ with the properly averaged stretching term,
\begin{equation}
  \dot{\overline{B}}_x = \overline{\vc{B}\cdot\nab u_x} \, .
\end{equation}
We can subsequently estimate $\alpha_{\rm dyn}$ from
\begin{equation}
  \alpha_{\rm dyn} \overline{B}_y = -\int \overline{\vc{B}\cdot\nab u_x} \de z
\end{equation}
by integrating the magnetic stretching term over $z$. This is of course only a
crude approximation of $\alpha_{\rm dyn}$ that ignores many of the
complications of analysing the evolution of the mean field component \citep[see
e.g.][]{Brandenburg2001,Brandenburg+etal2008}, but this method gives a good
order of magnitude estimate of the efficiency of creating radial field from the
large scale azimuthal field.

In the left panels of \Fig{f:alphadyn} we plot the integral $-\int
\overline{\vc{B}\cdot\nab u_x}\de z$ as a function of height over the midplane
and compare it to the azimuthal magnetic field $\overline{B}_y$. The
electromotoric force contribution of the magnetic stretching term is of
opposite sign at each side of the midplane. The correlation between
$\overline{B}_y$ and $-\int \overline{\vc{B}\cdot\nab u_x} \de z$ is shown in
the right panels of \Fig{f:alphadyn}, with orange (grey) symbols indicating
points above the midplane and blue (dark) symbols indicating points below the
midplane. Anticorrelation below the midplane and correlation above indicates
a positive $\alpha_{\rm dyn}$ above the midplane and a negative $\alpha_{\rm
dyn}$ below the midplane, of the order $\alpha_{\rm dyn}\approx0.023c_{\rm s}$
for run S3D\_256. The higher resolution run S3D\_512 gives $\alpha_{\rm
dyn}\approx0.032c_{\rm s}$. The fact that the helicity dynamo increases in
efficiency when going to higher resolution, even though both the collisional
hyper and shock resistivity coefficients are reduced, may indicate that the
outlined dynamo is a {\it fast dynamo}, although future simulations applying
resistivity on physical rather than on numerical grounds will be needed to
corroborate this point \citep[see e.g.][]{Hanasz+etal2002}.

Considering the creation of radial field by the Parker instability in a
galactic environment, \cite{HanaszLesch1993} predict $\alpha\simeq0.4\langle
v_\phi \rangle$, where the ``cyclonic velocity'' $\langle v_\phi \rangle$ of
\cite{Parker1979} is approximately $0.1 c_{\rm s}$ in our simulations. The
resulting dynamo coefficient $\alpha_{\rm dyn}\simeq0.04c_{\rm s}$ is quite
similar to what we find here from integrating the magnetic stretching term. In
the theoretical framework of \cite{Moffatt1978}, on the other hand, the value
of $\alpha_{\rm dyn}$ should be comparable to the correlation time of the
turbulence $\tau_{\rm corr}$ times the mean helicity, $\alpha_{\rm
dyn}\sim-(1/3)\tau_{\rm corr}\overline{(\nab \times \vc{u}')\cdot\vc{u}'}$, at
least in the limit of vanishing collisional resistivity. The helicity in our
simulations is positive below the midplane and negative above the midplane,
with an amplitude of $|(\nab\times\vc{u}')\cdot\vc{u}'|\sim1$. Thus the
inferred sign of $\alpha_{\rm dyn}$ fits well with the analytical theory, but
the absolute value of $\alpha_{\rm dyn}$ is at least ten times smaller than the
expectation based on a correlation time of order unity. This discrepancy may be
simply due to the fact that kinematic dynamo theory is not applicable to our
simulations, because the Lorentz force plays a significant role in determining
the evolution of the velocity field and the magnetic field.

\section{Summary and conclusions}
\label{s:conclusions}

In this paper we consider the evolution of strongly magnetised Keplerian
accretion discs. Our numerical experiments show that the hydromagnetic state of
the gas flow is very different from what is seen in zero net flux simulations.
The Parker instability leads to huge magnetic arcs rising several scale heights
from the disc midplane, and the magnetorotational instability in turn feeds
off the vertical fields and creates a highly turbulent flow, an interaction
that was predicted analytically by \cite{ToutPringle1992}. Although the flow is
stochastic and time fluctuating, we have identified an underlying dynamo
process that couples the vertically dependent mean radial and azimuthal
magnetic field components. As gas slides down inclined field lines, it obtains
a helical motion due to Coriolis forces, and thus the azimuthal field lines are
twisted in such a way as to create a mean electromotoric force in the direction
of the unperturbed field line -- a configuration prone to create radial field.
In turn the large scale radial field regenerates the azimuthal field through
Keplerian shear. Although Parker instability dominates the linear growth
phase, we have found evidence for magnerotational instability in the azimuthal
field as well. In the midplane of the disc, where the buoyancy is weak,
azimuthal MRI drives the initial evolution towards turbulence
\citep{FoglizzoTagger1994,FoglizzoTagger1995,TerquemPapaloizou1996}. These two
related instabilities, magnetorotational instability in the vertical fields
created by the Parker instability and magnetorotational swing instability in
the azimuthal fields, both rely on azimuthal flux confinement and can coexist in
the linear as well as in the non-linear state of transmagnetic accretion discs.

Such a path to accretion, based on the interaction of Parker and
magnetorotational instabilities, has at least two appealing traits. First of
all that the vertical fields that feed the magnetorotational instability are
created in a transparent way by the Parker instability. Zero magnetic flux
models must most likely rely on a small scale dynamo in order to create
vertical fields, and there is mounting evidence that such a dynamo would not
operate in the bulk part of accretion discs where the magnetic Prandtl number
is much lower than unity \citep{Schekochihin+etal2005,Fromang+etal2007}.
The second appealing result of our model is that the Maxwell and Reynolds
stresses are significant ($\alpha\approx0.1$). Such high accretion stresses
could solve the problem that observed accretion rates are often one or two
orders of magnitude higher than the accretion rates obtained in zero net flux
MRI simulations \citep{King+etal2007}.

The regeneration of azimuthal field by the shearing of an appropriate radial
field was seen in all our simulations that included Keplerian shear. As
magnetohydrostatic equilibrium is compromised by the Parker instability, gas
streams down along inclined field lines. Coriolis force diverts the gas to the
right, and a radial magnetic field is created as the azimuthal field is
subjected to shear-regions typically the size of the Parker instability.
Eventually magnetic reconnection leads to a coherent large scale radial
magnetic field. This dynamo was predicted by \cite{Parker1992} and subsequently
observed in the rigid rotation simulations of \cite{Hanasz+etal2002}. To our
knowledge we are the first to point out the relevance of Parker's fast galactic
dynamo to accretion discs and how it closes the accretion loop by replenishing
the azimuthal field that is lost by magnetic buoyancy.

The fine-tuned initial conditions with a purely azimuthal magnetic field and a
constant ratio of magnetic to thermal pressure may be questioned. However our
experiments with a combined azimuthal and vertical field shows that the Parker
instability is robust even if the azimuthal field coexists with a moderately
strong net vertical field, and that the additional vertical field component may
indeed increase the accretion rate further. 

Our results may also be relevant for star formation in the galactic centre.
Although there is currently no coherent accretion disc structure, the
population of young, massive stars in a disc-like structure close to the
galactic centre points towards the brief existence of an accretion disc some
million years ago
\citep{LevinBeloborodov2003,MilosavljevicLoeb2004,Nayakshin+etal2007,Alexander+etal2008}.
The disc was likely to be initially strongly magnetised, as indicated by the
current high magnetisation of the circumnuclear molecular ring. Hence the type
of MHD processes studied in this paper may be of central significance for the
disc dynamics. The presented simulations of transmagnetic ($\beta\sim1$) discs
argues that discs dominated by magnetic pressure $\beta \ll 1$ are
astrophysically viable. The existence of such discs was conjectured by
\cite{Pariev+etal2003} and their limitations and  observational consequences
were explored by \cite{BegelmanPringle2007}. A number of problems in accretion
disc theory are alleviated by the presence of super-equipartition magnetic
fields, among them is the long-standing issue of self-gravity and fragmentation
of AGN discs \citep{Goodman2003}. 

Future research into strongly magnetised accretion discs should also focus on
the self-consistent modelling of the
magnetisation of the material that feeds accretion discs in various
environments and on the evolution of magnetised disc coronae, in light of
effects such as reconnection, shearing of foot points, Ohmic heating and
radiative cooling that take place there.

\begin{acknowledgements}

Computer simulations were performed at the PIA cluster of the Max Planck
Institute for Astronomy and at the SARA Computing and Networking Service in
Amsterdam. We would like to thank Micha\l{} Hanasz, Steve Balbus, and Andrew
King for inspiring discussions. Axel Brandenburg, Jim Pringle and Richard
Alexander are thanked for commenting an early version of the manuscript.
We are grateful to Gordon Ogilvie and to the referee, Thierry Foglizzo,
for pointing out the relevance of the azimuthal field MRI for our simulations.
  
\end{acknowledgements}

\end{document}